\documentclass[notitlepage]{revtex4-1}

\usepackage{amsthm}
\usepackage{amsmath}
\usepackage{graphicx}
\usepackage{dcolumn}
\usepackage{bm}

\usepackage[utf8]{inputenc}
\usepackage[T1]{fontenc}
\usepackage{mathptmx}
\usepackage{multirow}
\usepackage{color}
\usepackage{float}
\usepackage[caption=false]{subfig}
\usepackage{caption}
\usepackage{hyperref}
\draft 

\begin{document}
	
	\title[Particle-laden turbulent Couette flow]{Dynamics  of particle-laden turbulent Couette flow. Part2: Modified fluctuating force model}
	
	\author{S.Ghosh}
	\altaffiliation[Also at ]{Department of Chemical Engineering,
		\\Indian Institute of Technology, Bombay}
	\email{swagnikg90@gmail.com}
	\author{P.S.Goswami}%
	\email{(for correspondence) psg@iitb.ac.in}
	\affiliation{Department of Chemical Engineering,
		\\Indian Institute of Technology, Bombay
		\\Powai, Mumbai-400-076, India
	}%
	
	\date{\today}

\begin{abstract}
Two-way coupled DNS simulation of particle-laden turbulent Couette-flow \citep{ghosh2022part1}, in the volume fraction regime $\phi>10^{-4}$, showed a discontinuous decrease of turbulence intensity beyond a critical volume fraction  $\phi_{cr}\sim7.875\times10^{-4}$. Due to the presence of high inertial particles, the drastic reduction of shear production of turbulence and in turn the reduction of viscous dissipation of turbulent kinetic energy is found to be the main cause for the discontinuous attenuation of turbulence.
 In this article, particle-phase statistics is explored. The two-way coupled DNS, reveal that the mean-square velocity profiles in cross-stream (y) and span-wise (z) directions are flat and increase with $\phi$ as the higher frequency of collision helps in transferring streamwise momentum to span-wise
and wall-normal directions. Whereas, streamwise fluctuations decrease and tend become flatter with increase in loading. In the regime with \textcolor{black}{$\phi>\phi_{cr}$}, the particle velocity fluctuations, which are generated by collisional re-distribution of momentum, drive the fluid phase velocity fluctuations. Additionally it is observed that one-way coupled DNS and Fluctuating Force Simulation (FFS) \citep{goswami2010particle} are capable to predict the particle phase statistics with reasonable accuracy in the regime $\phi<\phi_{cr}$ where wall-particle collision time and inter-particle collision time is lesser than viscous relaxation time of the particles. For, $\phi>\phi_{cr}$ a significant error in the prediction from one-way coupled DNS and FFS is observed. The reason for the deviation is found to be due to the limitation of FFS in capturing the turbulence attenuation and the change in mean fluid velocity profile. A modified FFS model (M-FFS) is successfully developed in this article with modified mean fluid velocity profile and zero-diffusivity. It is established here that the particle phase statistics of high inertial particles can also be predicted with less expensive decoupled M-FFS (modified FFS) model in place of two-way coupled DNS, in denser volume fraction regime, with an a priori knowledge of $\phi_{cr}$.
\end{abstract}

 \maketitle

\section{introduction}
In particle-laden turbulent flows, complex interaction between the carrier phase and the dispersed phase is present. Presence of wide-range of length and time-scale adds complexity to the problem. Consequently, simulating such flows from the first principle, namely DNS, becomes computationally expensive. Reduced order modelling strategies are especially significant in this context. The Mathematical framework for modelling the particle-laden turbulent flows can be of two types; Eulerian-Eulerian framework or Eulerian-Lagrangian framework. In Eulerian-Eulerian framework or two-fluid continuum modelling approach, both the fluid phase and the particle phase are considered to be inter-penetrating continuum. For high loading of the particles, when the integral fluid scale is higher than the inter-particle distance ($Kn<0.1$), this approach is relevant. Some seminal literature can be found in continuum modelling for the particle phase, where kinetic theory based Phase-space Probability Density Function (PDF) based approach is established. Evolution of these works can be understood from the review articles of \citet{reeks2014transport} and \citet{reeks2021development}. The main goal is
to find out a suitable master-equation to account for the statistical nature of the processes e.g. transport, mixing, and deposition of particles in turbulent flow. There are two formalisms of PDF approach. The first formalism, known as Kinetic Models (KM), refer to the probability density of a particle having a certain velocity $\mathbf{v}$ and position $\mathbf{x}$ at a given time ($t$) as the two variables \citep{reeks1991kinetic,reeks1992continuum,buyevich1971statistical,buyevich1972statistical1,buyevich1972statistical2}. In the second formalism of PDF based modelling is known as the
Generalised Langevin Models. This modelling involves velocity of the carrier phase flow local to a particle as an additional phase space variable, apart from particle position and velocity, of the probability density function \citep{haworth1985application,simonin1993eulerian,minier2014guidelines}. 
\\The Eulerian-Lagrangian framework involves tracking each of the particle of the dispersed phase individually and coupling those with the fluid phase descriptions in Eulerian grids. The DNS method is very accurate in predicting the dynamics of the both the phases as it uses governing equations from the first principle and fully resolves the fluid phase fluctuations from the largest scale to the smallest scale. However, due to the high computational cost of DNS, suitable reduced-oreder Eulerian-Lagrangian modelling is of importance. \citet{subramaniam2018towards} provided a very well-rounded review on the state-of-the-art modelling methodology in Eulerian-Lagrangian framework. If the particles are of sub-grid size and smaller than the Kolmogorov-scale, Point-particle approximation is used in DNS and thus resolving the flow across all the scales (PP-DNS). The point-particle Eulerian-Lagrangian modeling methodology involved suitable drag law model, improved interphase mass, momentum and energy transfer models and account for inter-particle interaction. Seminal work on deriving interphase momentum transfer term can be found in the works of \citep{garg2007accurate,garg2009numerically} .\citet{horwitz2016accurate} showed modification of Stokes drag considering the change in the fluid velocity field due to the effect of the reverse-drag of the neighbouring point-particles, in absence of any explicit inter-particle interaction. The effect of inter-particle interaction is incorporated in pairwise interaction extended point-particle or PIEP model by \citep{akiki2017pairwise}. The explicit account of inter-particle collision collision can not be done without any suitable deterministic collision model of the particles. The Fluctuating Force Simulation (FFS) developed by \citep{goswami2010particle} is of importance as it considered particle-particle and particle-wall collisions through event-driven simulation whereas the effect of fluctuating fluid velocity was modelled based on the Gaussian Random White Noise assumption which is applicable when the fluid integral time-scale is much less that the particle relaxation timescale. 

The turbulence modulation by particles \textcolor{black}{with high inertia ($St\sim367$) has been presented} thoroughly in \citet{ghosh2022part1}. \textcolor{black}{We have analyzed the} fluid phase velocity statistics and streamwise velocity and vorticity fluctuation field \textcolor{black}{using} two-way coupled DNS of \textcolor{black}{relatively dense} ($\phi=1.75\times10^{-4}-1.05\times10^{-3}$) particle-laden sheared turbulent suspension. The two-way fluid-particle interaction is such complex that even the nature of inter-particle collisions are found to have effect on the discontinuous turbulence transition. \textcolor{black}{In this regime particle-particle collision time is of the order or less than the viscous relaxation time of the particles.}
In the context of this intricate two-phase coupling, getting insights into the particle phase statistics in the above-mentioned volume-fraction regime is quite necessary \textcolor{black}{to device a fluid-particle decoupled modeling strategy to predict the particle phase dynamics. Fluctuating Force Simulation (FFS)  \citep{goswami2010particle1,goswami2010particle} is such a modelling methodology. In this work the dynamics of particle phase obtained from the model was investigated and validated by one-way coupled DNS simulations in turbulent Couette flow configuration.} 
\\ In this article, apart from analyzing the particle-phase statistics, investigation is made to explore the effect of reverse feedback force on the particle-phase statistics by analyzing the two-way coupled and one-way coupled DNS \textcolor{black}{for wide range of volume fraction}. Lastly, the applicability of the FFS model \textcolor{black}{is demonstrated for this range of volume fraction}. 
\section{Simulation Methodology}
The system geometry and dimensions and all the necessary parameters for the two-way coupled DNS simulation is discussed in \citet{ghosh2022part1}. The two-way coupled DNS results are compared with data obtained by one-way coupled DNS and Fluctuating Force Simulation. A brief recap of the FFS is done in the subsection \ref{sec:FFS}. The collision rule remains same for DNS and FFS and is discussed in subsection \ref{sec:coll_rule}.
\subsection{The Fluctuating Force Simulation (FFS) model}
\label{sec:FFS}
\citet{goswami2010particle} obtained the statistics of particle phase using fluid velocity fluctuations modeled as white noise and embedding this random forcing on the particle phase through a hard sphere molecular dynamics type simulation. The strength of the noise was calculated from  direct numerical simulation for unladen flow. From the analysis of fluid phase velocity and and particle phase acceleration distribution it was observed  that the fluctuating acceleration distribution statistics of the particle phase is identical to the velocity distribution statistics of fluid phase with a scale factor of particle relaxation time. With this observation the authors calculated the velocity space diffusion coefficient from the autocorrelation of the fluid phase velocity fluctuation. 
Where $D_{ij}$ is velocity space diffusion coefficient and can be expressed as 
\begin{equation}
	\label{eq2.4}
	D_{ij}=\int_{0}^{\infty}dt^{'}\langle a_i^{'}(t^{'})a_j^{'}(0)\rangle=\frac{1}{\tau_v^2}\int_{0}^{\infty}dt^{'}\langle u_i^{'}(t^{'})u_j^{'}(0)\rangle
\end{equation}
The above relation is valid when acceleration in the particle phase acceleration $a'$ is solely caused by velocity fluctuation in fluid phase $u'$. In the above relation the time scale of velocity fluctuation in the fluid phase is less than viscous relaxation time so that the effect of fluid velocity fluctuations can be modeled as Gaussian white noise. The diffusion tensor $D_{ij}$ is symmetric and anisotropic with each element having dimension $length^2/time^3$ . Among the off-diagonal components of $D_{ij}$, $D_{xz}$ and $D_{yz}$ are zero for the shear-flow taking place in x-y plane along x-direction, as probability of velocity fluctuation along z axis is even function of velocity. For a single particle the Langevin type equation equivalent to Fokker-Planck equation given in \citep{goswami2010particle} can be written as
\begin{equation}
	\label{eq2.5}
	\frac{dv_i}{dt}=-\frac{v_i-\bar{u_i}(\bf{x}_p)}{\tau_v}+F_i(t)
\end{equation}
Here, $v_i$ represents the velocity of the particle and $\bar{u_i}(\bf{x}_p)$ represents fluid mean velocity along i-th direction interpolated at particle location. The first term in the right hand side thus represents the drag force on the particle due to mean fluid flow. The second term $F_i(t)$  is the fluctuating force acting on the particle due to the velocity fluctuations in the fluid phase and this term is modeled as Gaussian white noise with zero mean and the second moment calculated as follows:
\begin{equation}
	\label{eq2.6}
	\langle F_i(t)F_j(t)\rangle=2D_{ij}\delta(t-t^{\prime})
\end{equation}
In the above equation $D_{ij}$ is the diffusion tensor as defined earlier.

Equation \ref{eq2.5} with the second moment of the force described by equation \ref{eq2.6} and additional consideration of particle-particle collisions is used for  'fluctuating force' simulation. 
Equation \ref{eq2.5} can be represented as a difference equation as follows:
\begin{equation}
	\label{eq2.7}
	v_i(t+\Delta t)-v_i(t)=\frac{v_i-\bar{u_i}}{\tau_v}\,\Delta t+F_i(t)\Delta t
\end{equation}
The time interval $\Delta t$  is smaller than the fluid integral time. In generating the magnitude of the random force, three independent random normal deviates with zero mean and unit variance $\zeta_1$, $\zeta_2$, $\zeta_3$ are generated and the components of the
fluctuating force can be represented as
\begin{equation}
	\label{eq2.11}
	F_x=\frac{\sqrt{2D_{xx}}}{\sqrt{\Delta t}}\zeta_1
\end{equation}

\begin{equation}
	\label{eq2.12}
	F_y=\frac{\sqrt{2D_{yy}}}{\sqrt{\Delta t}}\left[\frac{D_{xy}}{\sqrt{D_{xx}D_{yy}}}\zeta_1+ \sqrt{1-\frac{D_{xy}^2}{D_{xx}D_{yy}}}\zeta_2 \right]
\end{equation}

\begin{equation}
	\label{eq2.13}
	F_z=\frac{\sqrt{2D_{zz}}}{\sqrt{\Delta t}}\zeta_3
\end{equation}
The above expressions can be derived from equation \ref{eq2.6} replacing $\delta (t-t^{\prime})$ with
$\frac{1}{{\Delta}t}$. 
\subsection{The collision rule}
\label{sec:coll_rule}
In order to simulate the particle motion realistically, the effect of the Fluctuating force and fluid drag force is not sufficient unless particle-particle interactions and particle-wall interactions are properly modeled. Even if the particles are assumed to be hard particles, to address the particle dynamics for any system with practical importance, we need to  incorporate inelasticity and roughness factor in the collisions model. The inelastic nature can be accounted by introducing co-efficient of restitution $e$ that signifies the loss of translational kinetic energy in two particle collision (\textcolor{black} {\citet{bose2004velocity}}). This co-efficient of restitution $e$ ranges between 0 and 1. The case $e=1$ represents perfectly elastic collision whereas lower values of $e$ indicates departure from the elastic behaviour.

Let us  consider  $\mathbf{v_1},m_1$  be the linear velocity and mass respectively of the first colliding particle. The same notations with a subscript 2 denote the same quantities for the second particle of the colliding pair. Vector $\mathbf{k}$ represents the unit vector along the line joining centres from particle 1 to 2. The velocity of the centre of mass of the 1st particle as observed from the centre of mass of the 2nd particle is written as:
\begin{math}
\mathbf{v_{12}}=\mathbf{v_{1}}-\mathbf{v_{2}}
\end{math}
\\Let us define $\eta_1=1+e$ and $m_{12}=\frac{m_1m_2}{m_1+m_2}$

If $v_1^{'}$ and $v_2^{'}$ be the post collision velocities of the particles, then the final changes in linear velocities of the colliding bodies after the collision can be written as:
\begin{align}
	\label{eq2.27}
	m_1(\mathbf{v_1}^{'}-\mathbf{v_1})=m_2(\mathbf{v_2}-\mathbf{v_2}^{'})=-m_{12}\eta_1(\mathbf{k}.\mathbf{v_{12}})\mathbf{k}
\end{align}
Hence, at the instant between two successive noise-injections due to the fluctuating force the equation of motion of colliding particles subjected to inter-particle collision or wall-particle collision are captured through equation \ref{eq2.27}.
\section{Particle Phase Statistics for Elastic Collisions}
\label{sec:Particle Phase Statistics ideal}

\begin{figure*}[!]
			\includegraphics[width=1.0\textwidth]{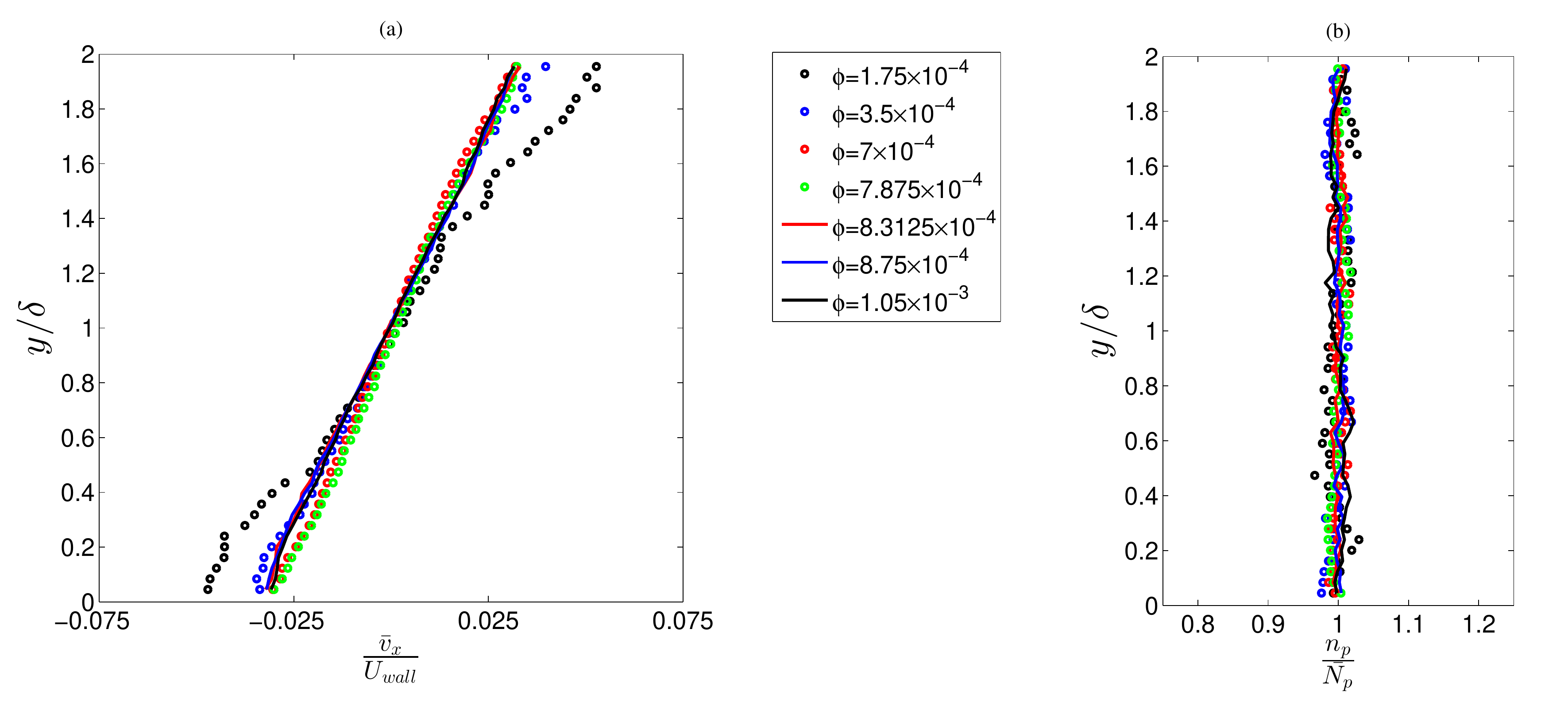}
			\caption{Effect of particle volume fraction on (a) mean velocity, and (b) particle number concentration } 
			\label{fig:particle_phase_stat_1}
		\end{figure*}

\begin{figure*}[!]
			\includegraphics[width=1.0\textwidth]{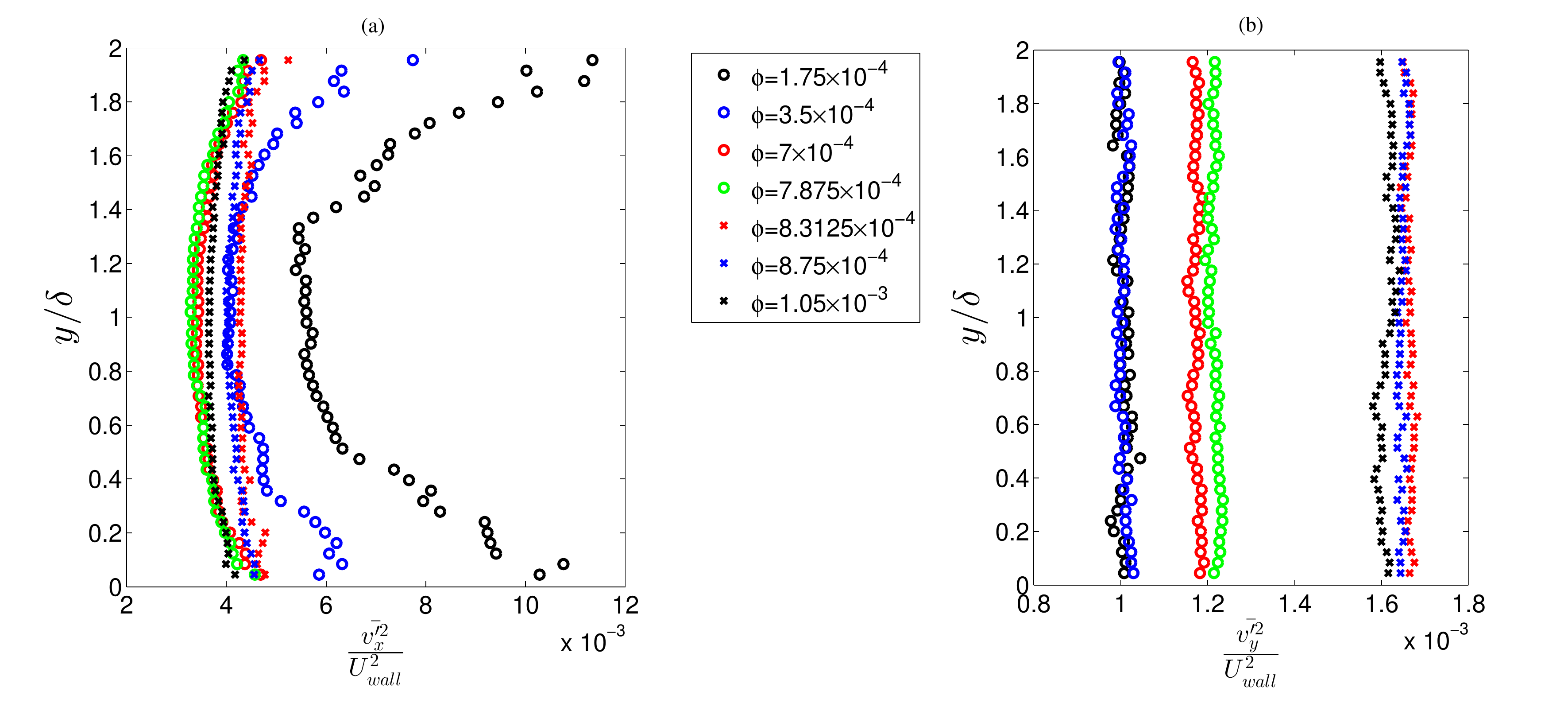}
			\caption{Effect of particle volume fraction on (a) streamwise and (b) cross-stream mean square velocity of the particle phase}
			\label{fig:particle_phase_stat_2} 
		\end{figure*}

\begin{figure*}[!]
			\includegraphics[width=1.0\textwidth]{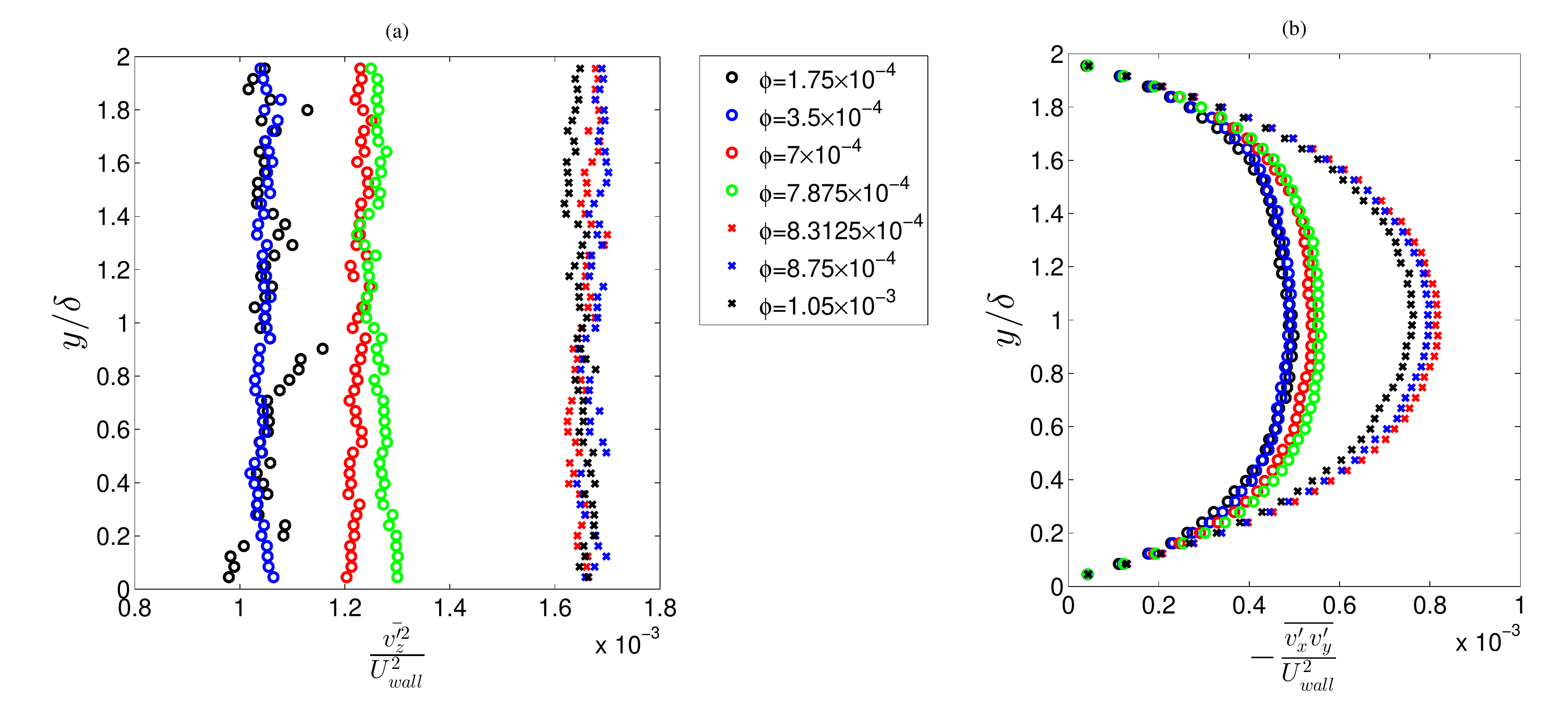}
			\caption{Effect of particle volume fraction on (a) span-wise mean square velocity, (b) streamwise and cross-stream velocity cross-correlation of the particle phase} 
			\label{fig:particle_phase_stat_3}
		\end{figure*}

\begin{figure}[!]
 	\includegraphics[width=0.6\linewidth]{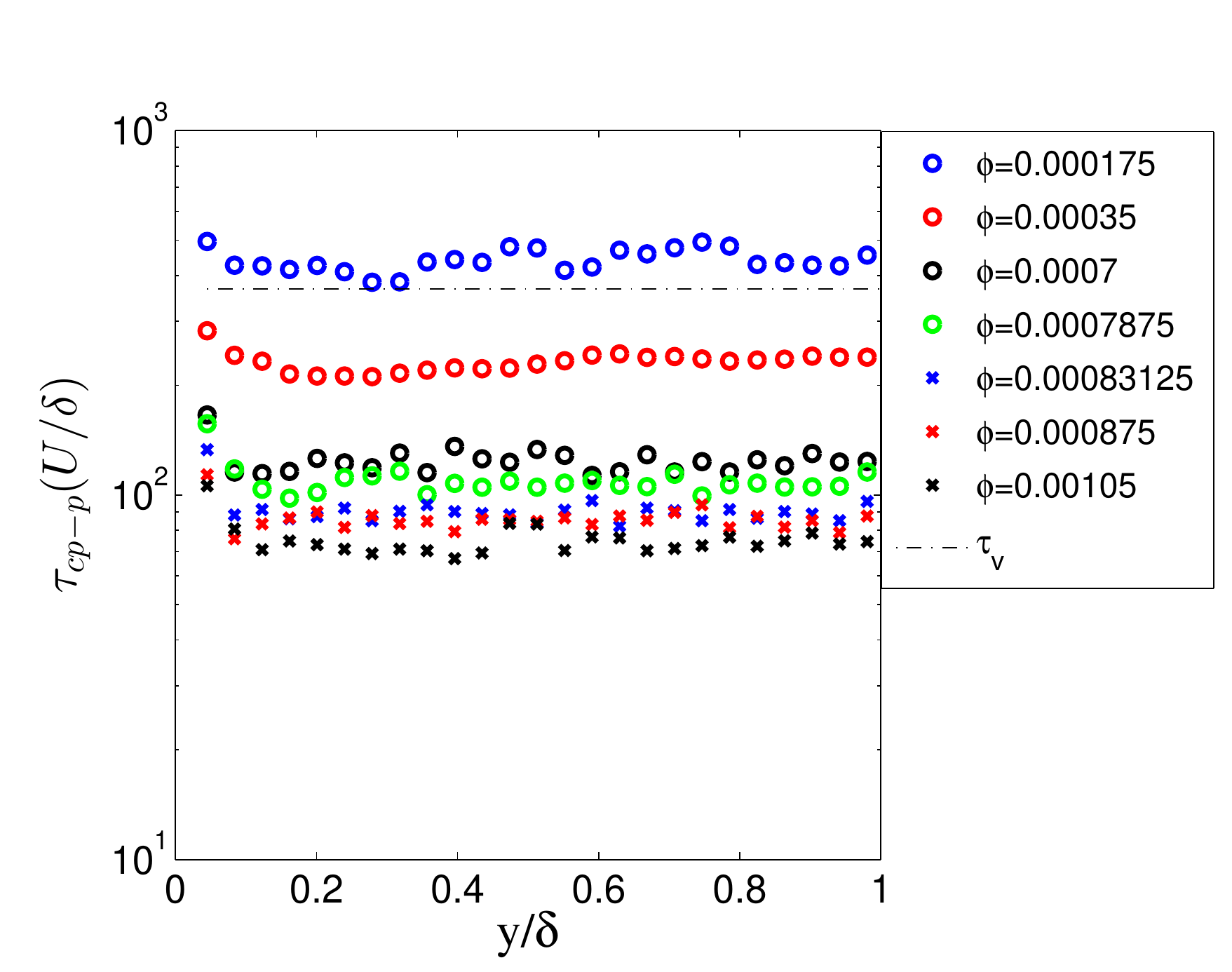}
 	\caption*{(a)}
 	\includegraphics[width=0.6\linewidth]{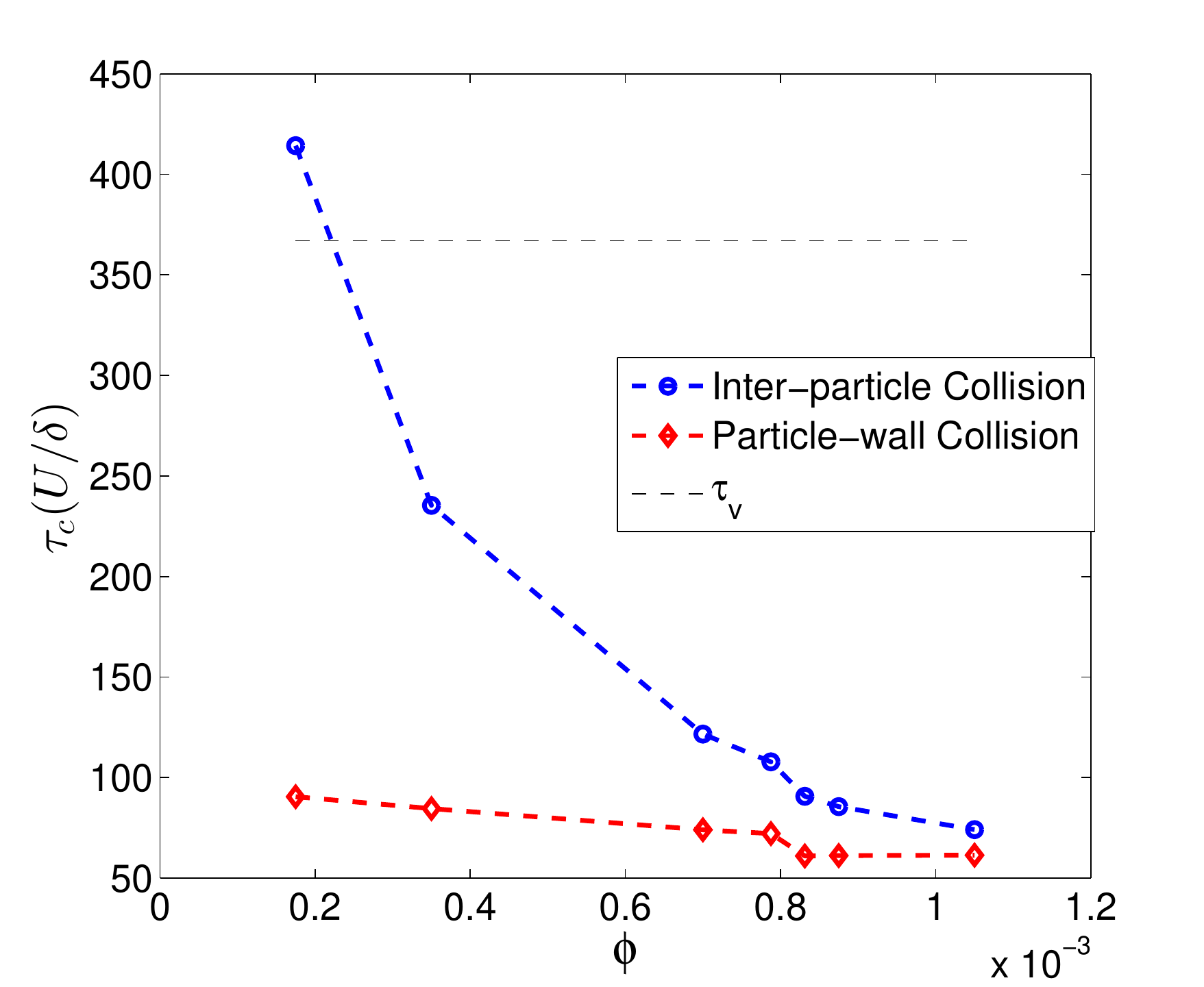}
 	\caption*{(b)}
 	\caption{Collision statistics: (a) profile of local inter-particle collision time ($\tau_{cp-p}$) across the Couette-flow $y/\delta$ for different $\phi$, (b) variation of overall collision time ($\tau_c$) for both inter-particle and wall-particle collisions with $\phi$}
 	\label{fig:coll_stat}
\end{figure}

The investigation of particle phase statistics are especially necessary in order to relate \textcolor{black}{it to} the fluid phase turbulence modulation and \textcolor{black}{to obtain a picture beyond critical} volume fraction ($\phi_{cr}=7.875\times10^{-4}$). The particle phase statistics are \textcolor{black}{presented} in figures \ref{fig:particle_phase_stat_1} to \ref{fig:particle_phase_stat_3}.
 Figure \ref{fig:particle_phase_stat_1} shows the mean particle velocity \textcolor{black}{(fig \ref{fig:particle_phase_stat_1}(a)) along with the particle number concentration profile (fig \ref{fig:particle_phase_stat_1}(b)) in} the Couette-flow. Mean velocity of the particles remain higher near the wall at the lowest volume fraction; although for the rest of $\phi$ values, the effect of volume loading is \textcolor{black}{insignificant} even above critical volume fraction. The particle concentration profile remains \textcolor{black}{almost constant for all the volume fractions reported here.} 
 In figures \ref{fig:particle_phase_stat_2}-\ref{fig:particle_phase_stat_3} it is observed that unlike $\overline{v_y^{\prime 2}}$ and $\overline{v_z^{\prime 2}}$, $\overline{v_x^{\prime 2}}$ has a variation with respect to cross-stream position ($y/\delta$); being higher at the walls and lowest at the center of the Couette-flow (Fig. \ref{fig:particle_phase_stat_2}(a)). With the increase in $\phi$ value, $v_x^{\prime 2}$ decreases and the profile becomes flatter. Figures \ref{fig:particle_phase_stat_2}(b) and \ref{fig:particle_phase_stat_3}(a) show that the profiles of mean square velocities $\overline{v_y^{\prime 2}}$ and $\overline{v_z^{\prime 2}}$ respectively are mostly constant across the channel. As the walls are perfectly smooth and wall-particle collisions are perfectly elastic, they do not act as any source of transverse momentum. \textcolor{black}{The particle concentration is also constant in the wall normal direction.} Consequently, 
 $\overline{v_y^{\prime 2}}$ becomes \textcolor{black}{constant in the wall-normal direction}. These fluctuations increase with $\phi$ upto the critical volume fraction, $\phi_{cr}=7.875\times10^{-4}$. After \textcolor{black}{that stream wise} mean square velocity fluctuations show marginal decrease with increase in $\phi$. Similar observation is noted for the particle velocity cross correlation $-\overline{v_x^{\prime}v_y^{\prime}}$ (fig \ref{fig:particle_phase_stat_3}(b)), which unlike the other moments have highest magnitude at the center of the Couette-flow and the lowest at the walls. \textcolor{black}{\citet{muramulla2020disruption} have reported} similar trends in particle mean square velocities before and after the discontinuous transitions in their two-way coupled channel-flow simulation.
\\ The inter-particle collision time across the \textcolor{black}{wall-normal direction are shown in Fig. \ref{fig:coll_stat}(a). Collision time is found to be slightly higher at the walls.} Except for the lowest volume fraction, the inter-particle collision time ($\tau_{cp-p}$) remains lower than the particle-relaxation time ($\tau_v$) across the cross-stream position indicating that particles undergo successive collision with another particle well before \textcolor{black}{it relaxes to local velocity}. Figure \ref{fig:coll_stat}(b)  shows  the overall collision time (i.e. the collision times calculated over the entire simulation box) for both inter-particle and wall-particle collisions. \textcolor{black}{Collision time decreases as expected} with increase in volume fraction. Also it is observed that, except for $\phi=1.75\times10^{-4}$, inter-particle collision time is less than $\tau_v$ and for all volume fraction overall particle-wall collision time is \textcolor{black}{much lower} than both inter-particle collision time and $\tau_v$. This indicates that the dynamics of the particles are mainly governed by wall-particle and inter-particle collision \textcolor{black}{over the} viscous relaxation when $\phi>1.75\times10^{-4}$. Due to the high volume fraction of the heavy inertial particles, the particles distribute uniformly across the channel. 
The decrease in particle streamwise mean square velocity with increase in volume fraction, is due to the \textcolor{black}{attenuation of} fluid streamwise fluctuation.
\textcolor{black}{At higher volume loading the profiles become flatter due to the increase in collisional transfer of momentum in the cross-stream direction. Streamwise particle velocity fluctuation decreases up to critical volume loading, then there is a slight increase in velocity fluctuation when volume fraction is increased beyond the $\phi_{cr}$.}
The mean-square velocity profiles in cross-stream and spanwise direction are flat and increase with volume fraction \textcolor{black}{as the higher frequency of collision helps in transferring streamwise momentum to span-wise and wall-normal directions.}
It is observed from the two-way coupled DNS simulation that above $\phi_{cr}$, the mean square fluid phase velocity fluctuations decrease so drastically that the particle phase mean square velocity fluctuations, 
become two orders of magnitude larger than the fluid phase velocity mean square values. \textcolor{black}{It is to be noted that at a volume fraction lower than the $\phi_{cr}$, fluid phase fluctuation is higher than the particle phase fluctuation}. Hence in the regime with \textcolor{black}{$\phi>\phi_{cr}$}, the particle velocity fluctuations, which are generated by collisional re-distribution of momentum, drive the fluid phase velocity fluctuations.
\section{Extent of Deviation of Particle Phase Statistics Predicted by One-way Coupling}
\label{sec:one-way_two_way}
Discussion in the previous section reveals the effect of collisional redistribution of momentum on particle-phase statistics \textcolor{black}{for a wide range of particle volume fraction.} 
\begin{figure*}[!]
			\includegraphics[width=1.0\linewidth]{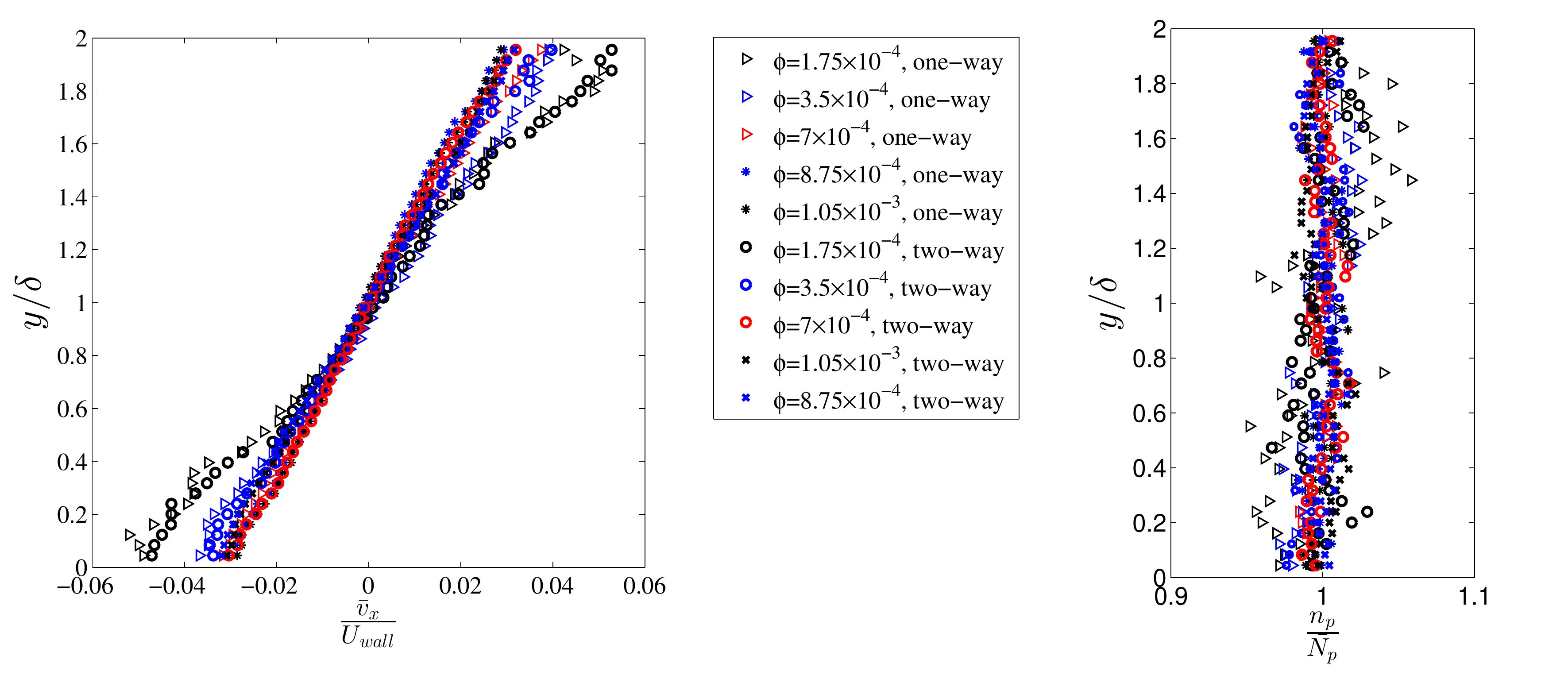}
			\caption{Effect of reverse feedback force at different particle volume fraction on (a) mean velocity, and (b) particle number concentration} 
			\label{fig:particle_phase_stat1_1way_2way}
\end{figure*}
\begin{figure*}[!]
			\includegraphics[width=1.0\linewidth]{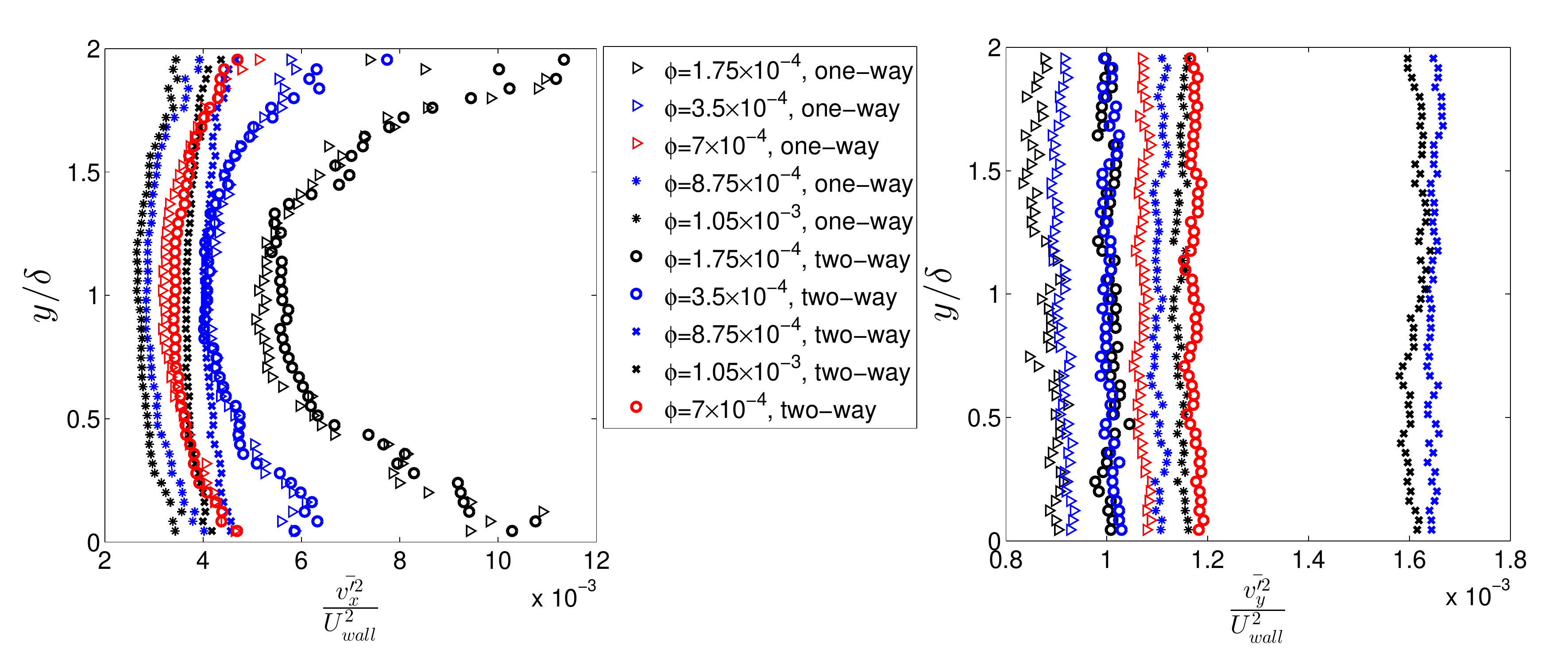}
			\caption{Effect of reverse feedback force at different particle volume fraction on  (a) streamwise and (b) cross-stream mean square velocity of the particle phase}
			\label{fig:particle_phase_stat2_1way_2way} 
\end{figure*} 
\begin{figure*}[!]
			\includegraphics[width=1.0\linewidth]{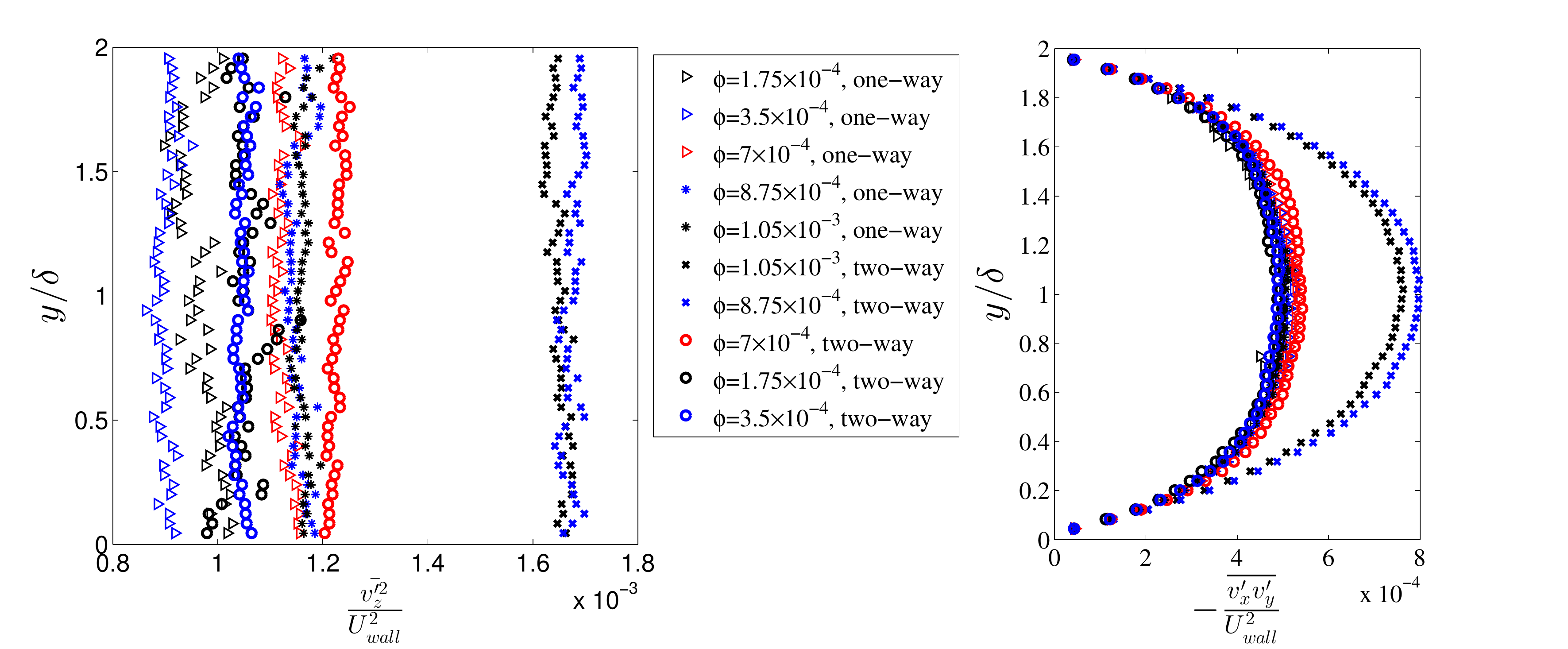}
			\caption{Effect of reverse feedback force at different particle volume fraction on (a) span-wise mean square velocity, (b) streamwise and cross-stream velocity cross-correlation of the particle phase} 
			\label{fig:particle_phase_stat3_1way_2way}
\end{figure*}
\textcolor{black}{Due to high frequency of collisions, particle phase fluctuations increase even at a higher volume fractions where fluid turbulence collapses. Therefore, the usual question arises whether an one-way coupled DNS or any model developed based on the turbulent diffusivity calculated from the one way coupled DNS can successfully predict the particle dynamics in the limit of high Stokes number. To answer this, we have compared the particle velocity statistics obtained from one-way and two-way coupled simulations. Here, we report the statistics for a volume fraction ($\phi$) range of $1.75\times10^{-4}$ to $1.05\times10^{-3}$. Figure \ref{fig:particle_phase_stat1_1way_2way} (a) and (b) show that the predictions of mean velocity and concentration profile from one-way coupled case varies marginally from that of the two-way coupled simulation.}
Figure \ref{fig:particle_phase_stat2_1way_2way} and \ref{fig:particle_phase_stat3_1way_2way} represents the second moment of the particle phase velocities. The \textcolor{black}{difference of} second moments \textcolor{black}{predicted by one-way and two-way coupled simulations} is prominent for volume fractions greater than $\phi_{cr}$. Above $\phi_{cr}$ absence of reverse-feedback force (\textcolor{black}{one-way coupling) underpredicts} the streamwise, cross-stream and span-wise mean square velocities by approximately $25\%-30\%$. This \textcolor{black}{underprediction} is about $30-40\%$ for \textcolor{black}{$\overline{v_x'v_y'}$}. In one-way coupled DNS, the fluid phase velocity remains same as the unladen flow. 
\textcolor{black}{Therefore, any modification in fluid mean velocity or fluid velocity fluctuation which may alter the particle phase velocity fluctuations at higher particle volume loading can not be captured using one-way coupling. However, we observe that at lower volume fraction $\phi<\phi_{cr}$, the second moments of the particle velocity fluctuation predicted by one-way coupling differs marginally ($<10\%$) from the prediction of two-way coupling. Such an observation suggests that any model developed based on the unladen fluid phase fluctuation may successfully predict the particle phase dynamics.\\
The pdf of particle velocity fluctuation} are shown in figures \ref{fig:oneway_two_way_fvx} and \ref{fig:oneway_two_way_fvy} for two different volume fractions $\phi=7\times10^{-4}$ (lower than $\phi_{cr}$) and $\phi=1.05\times10^{-3}$ (greater than $\phi_{cr}$). Here the fluctuating velocity components $v_i'$ are denoted as $v_i$. The Effect of reverse force is not prominent in case of $f(v_x)$ even when $\phi=1.05\times10^{-3}$. However, the effect of reverse force is observed for $f(v_y)$ when $\phi=1.05\times10^{-3}$. The tails of the pdfs are found to be elongated in presence of reverse force. This is consistent with the trends in $\overline{v_y'^2}$ as observed in \ref{fig:particle_phase_stat2_1way_2way}(b).                
\begin{figure*}
    \centering
    \includegraphics[width=1.0\linewidth]{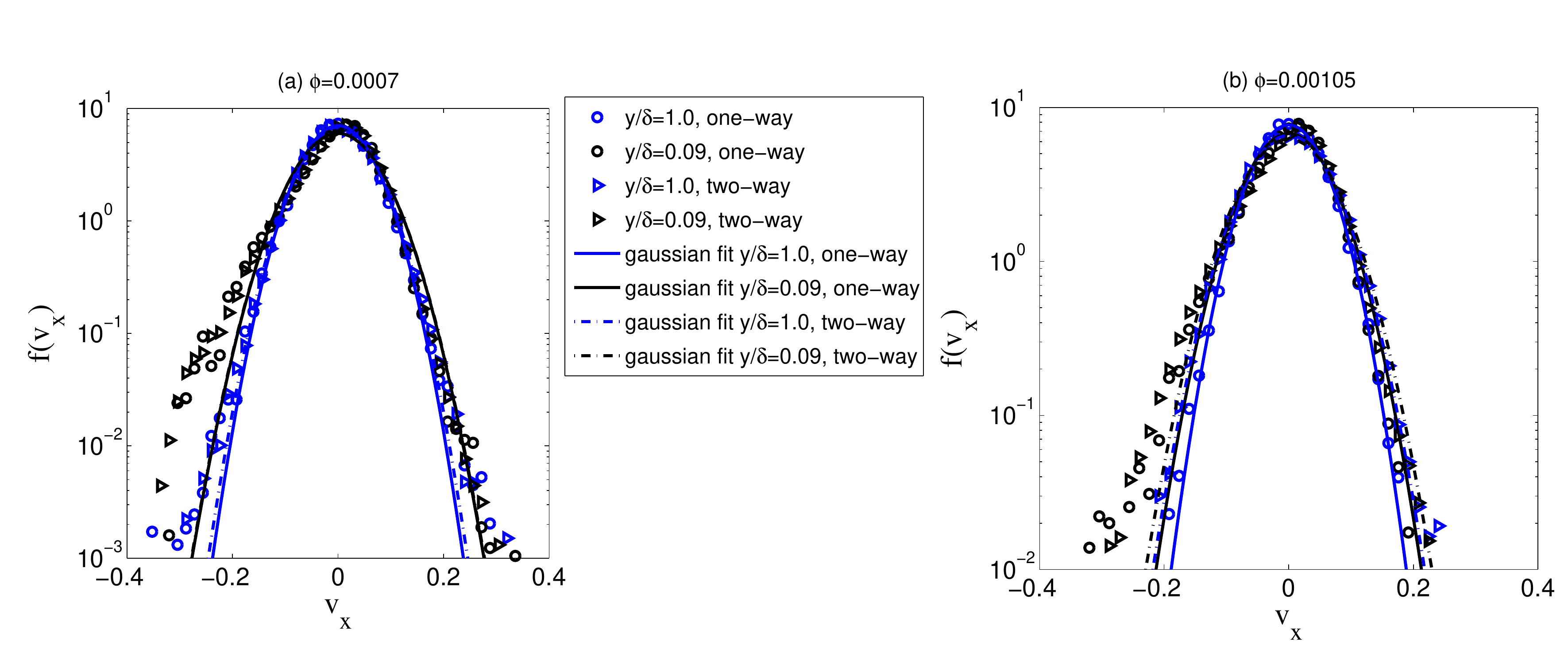}
    \caption{Effect of reverse feedback force on particle streamwise velocity distribution function $f(v_x)$ for (a) $\phi=7\times10^{-4}<\phi_{cr}$ and (b) $\phi=1.05\times10^{-3}>\phi_{cr}$}
    \label{fig:oneway_two_way_fvx}
\end{figure*}
\begin{figure*}
    \includegraphics[width=1.0\linewidth]{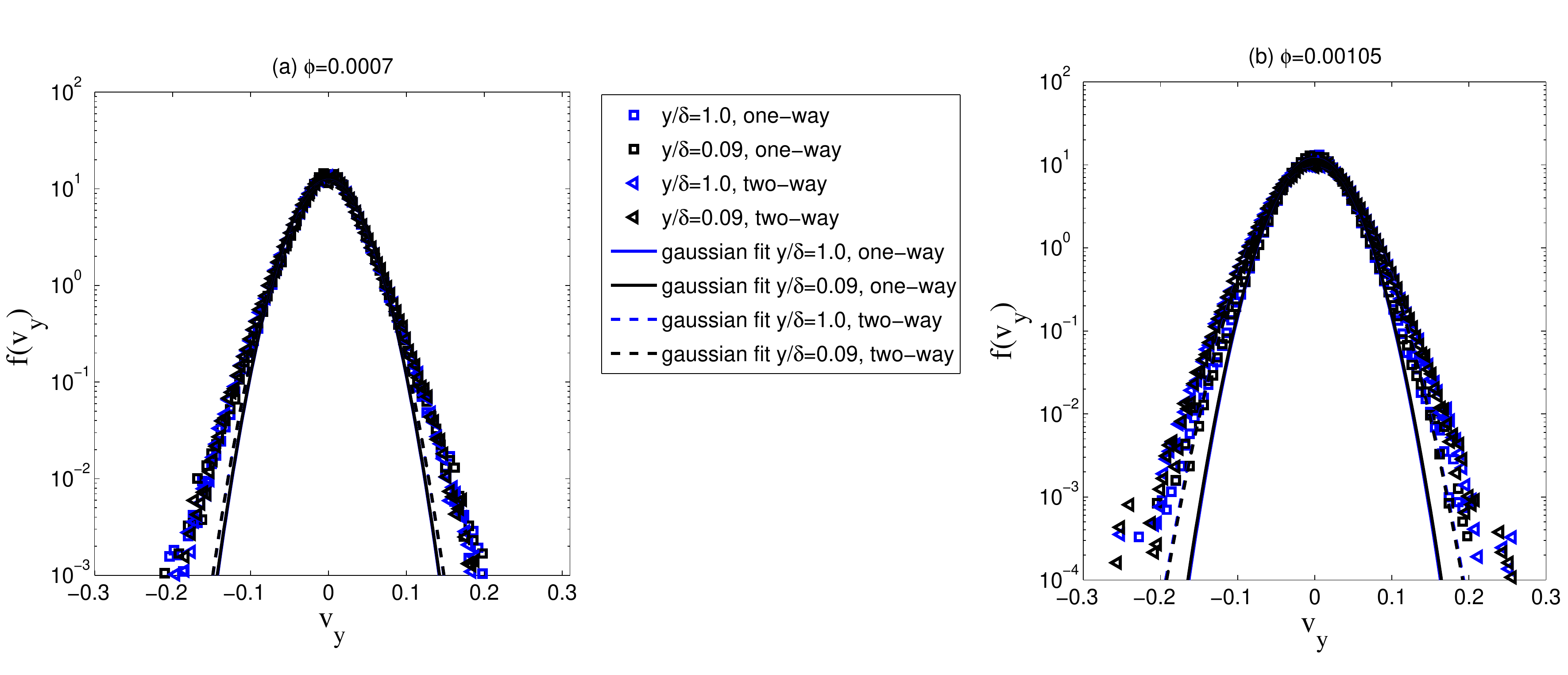}
      \caption{Effect of reverse feedback force on particle cross-stream velocity distribution function $f(v_y)$ for (a) $\phi=7\times10^{-4}<\phi_{cr}$ and (b) $\phi=1.05\times10^{-3}>\phi_{cr}$}
    \label{fig:oneway_two_way_fvy}
\end{figure*}

\begin{figure*}[!]
 \includegraphics[width=0.8\linewidth]{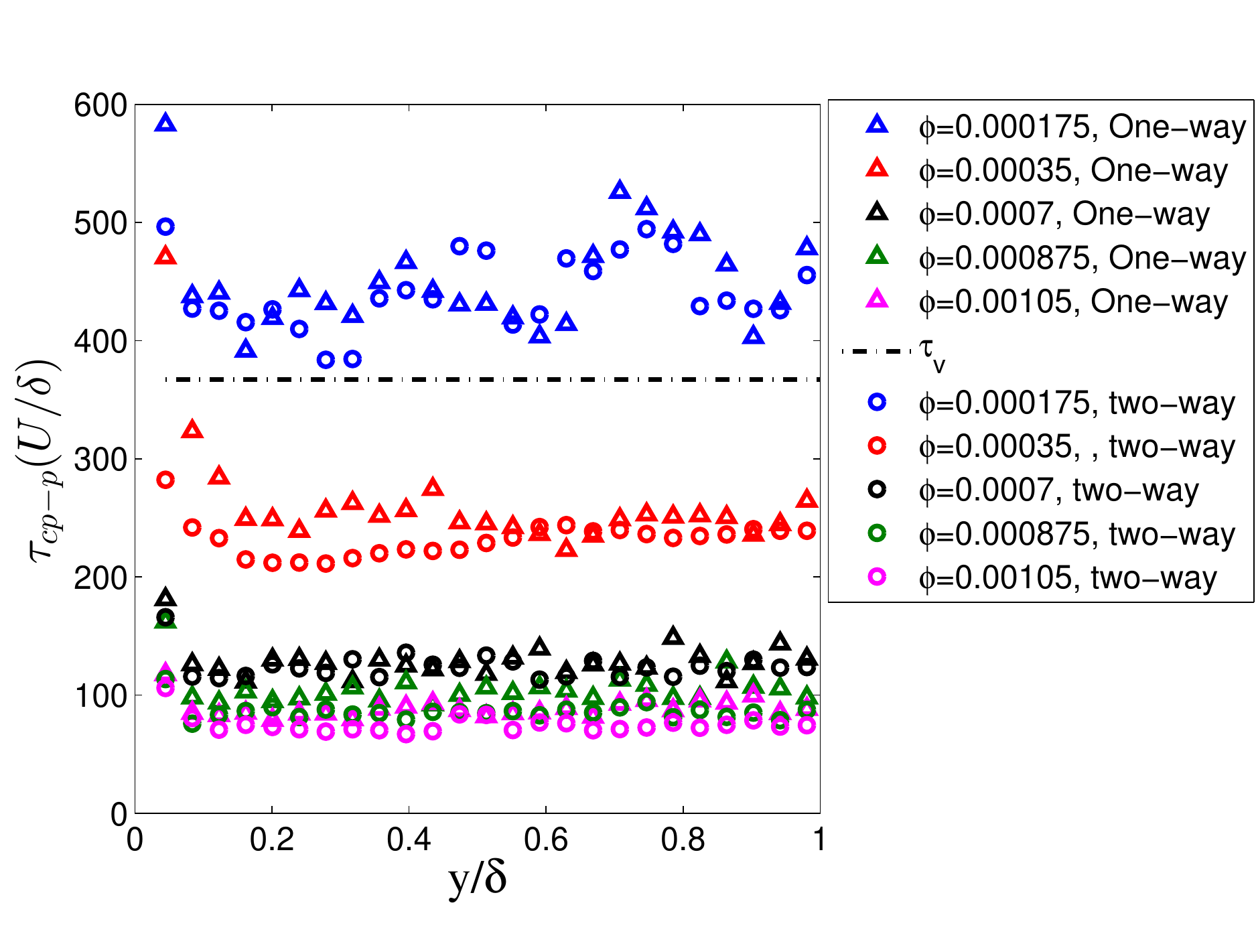}
 	\caption*{(a)}    
\includegraphics[width=0.6\linewidth]{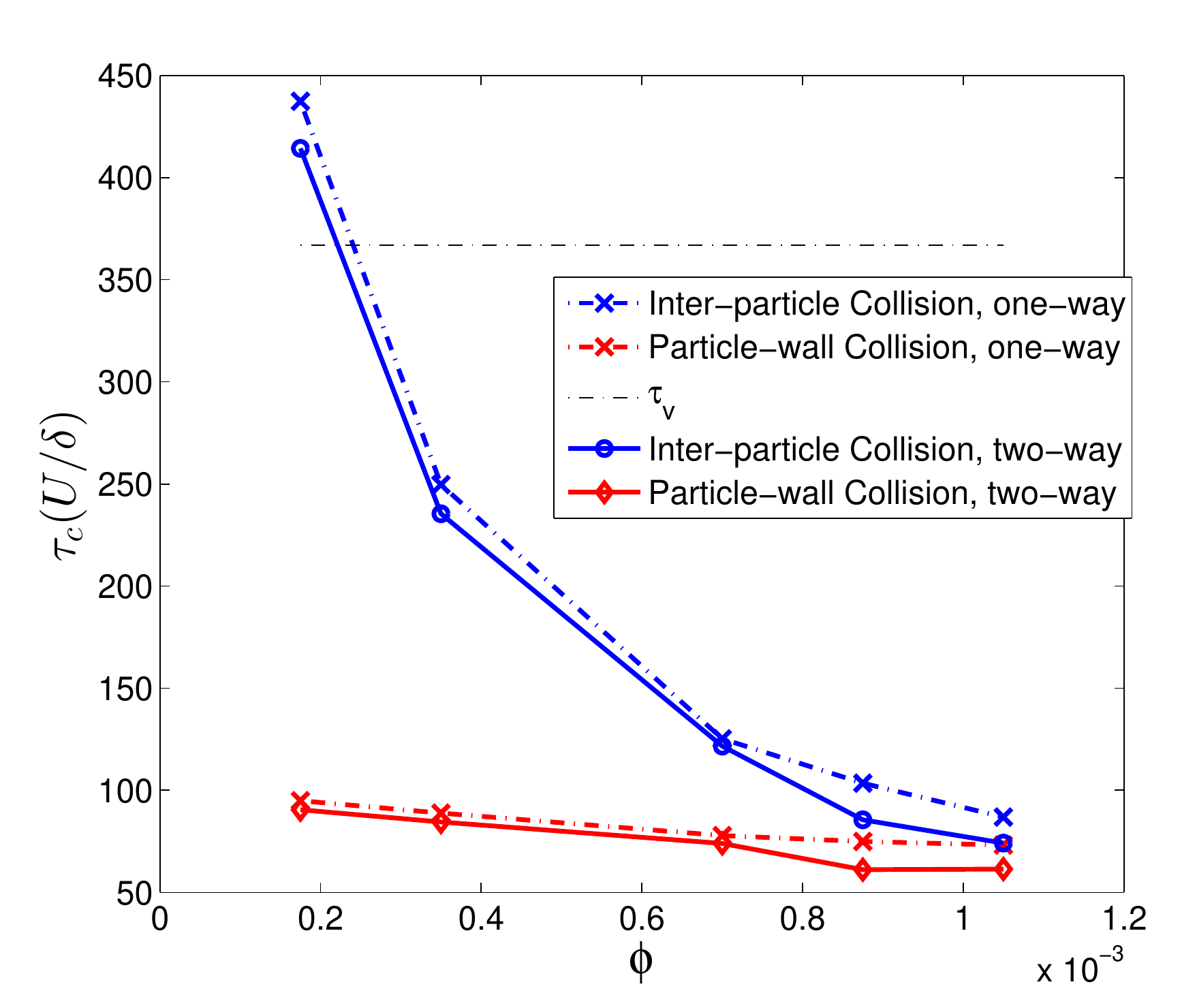}
 	\caption*{(b)}     
\caption{Effect of reverse feedback force on (a) profile of inter-particle collision time $\tau_{cp-p}$, (b) variation of average inter-particle and wall-particle collision time with $\phi$} 
\label{fig:coll_one_way_two_way}
\end{figure*}

\textcolor{black}{Besides obtaining the velocity statistics, we have also computed the particle-particle and particle-wall collision times.} The profile of inter-particle collision time is shown in figure \ref{fig:coll_one_way_two_way}(a). In presence of reverse force, the decrease in inter-particle collision time is visible at higher volume fractions greater than $\phi_{cr}$. Figure \ref{fig:coll_one_way_two_way}(b) shows that the average inter-particle collision time and wall-particle collision time decreases significantly for denser suspensions with $\phi>\phi_{cr}$. These plots show that $\tau_{cp-w} < \tau_{cp-p} < \tau_v $ except for $\phi=1.75\times10^{-4}$ where $\tau_{cp-w} < \tau_v< \tau_{cp-p} $. Hence collisional dynamics, especially  wall-particle collisions, play a more important role in the particle phase dynamics. \textcolor{black}{Figure \ref{fig:coll_one_way_two_way}(b) depicts that the collision frequency is underpredicted by a maximum of about $21\%$ for wall-particle collisions and about $16\%$ for particle-particle collisions by one-way coupling when $\phi>\phi_{cr}$. The discrepancy may originate due to the inability of the one-way coupling to capture the modification of fluid phase mean velocity and velocity fluctuations.\\
To summarize, in the limit of high Stokes number, at $\phi<\phi_{cr}$ one-way coupling is capable to predict the particle phase statistics with reasonable accuracy, but at $\phi>\phi_{cr}$ there happens a significant error in the prediction. In the following section we demonstrate the applicability of FFS \citep{goswami2010particle} and discuss about the required modifications to predict the particle dynamics for wide range of solid volume fraction.}  
\section{Applicability of FFS Model}
\label{App_FFS}
\subsection{\textcolor{black}{Comparison of FFS prediction with DNS results}}
\label{one-way_two_way_FFS}
The discussion in the previous section (section \ref{sec:one-way_two_way}) shows that \textcolor{black}{in the limit of high St}, for the suspensions with volume fractions less than $\phi_{cr}$, one-way coupled DNS \textcolor{black}{may} be used to predict the particle phase velocity statistics with little inaccuracy ($<10\%$). \textcolor{black}{In this context it is very much} relevant to investigate the applicability of FFS model, \textcolor{black}{described} in \citet{goswami2010particle}, for denser turbulent suspensions with volume fraction less than $\phi_{cr}$.
Comparative study is done among one-way coupled DNS, two-way coupled DNS and FFS results for three different volume fractions. Two Runs are made for $\phi=3.5\times10^{-4}$ and $\phi=7\times10^{-4}$ representing the regime \textcolor{black}{with $\phi<\phi_{cr}$} and for $\phi=8.75\times10^{-4}$ representing the regime \textcolor{black}{with $\phi>\phi_{cr}$}. It is to be mentioned that all the inter-particle and wall-particle collisions are considered to be perfectly elastic. \textcolor{black}{In this section we mainly focus on the accuracy of FFS prediction for particle phase mean and r.m.s. velocities.}
\begin{figure*}[!]
			\includegraphics[width=1.0\linewidth]{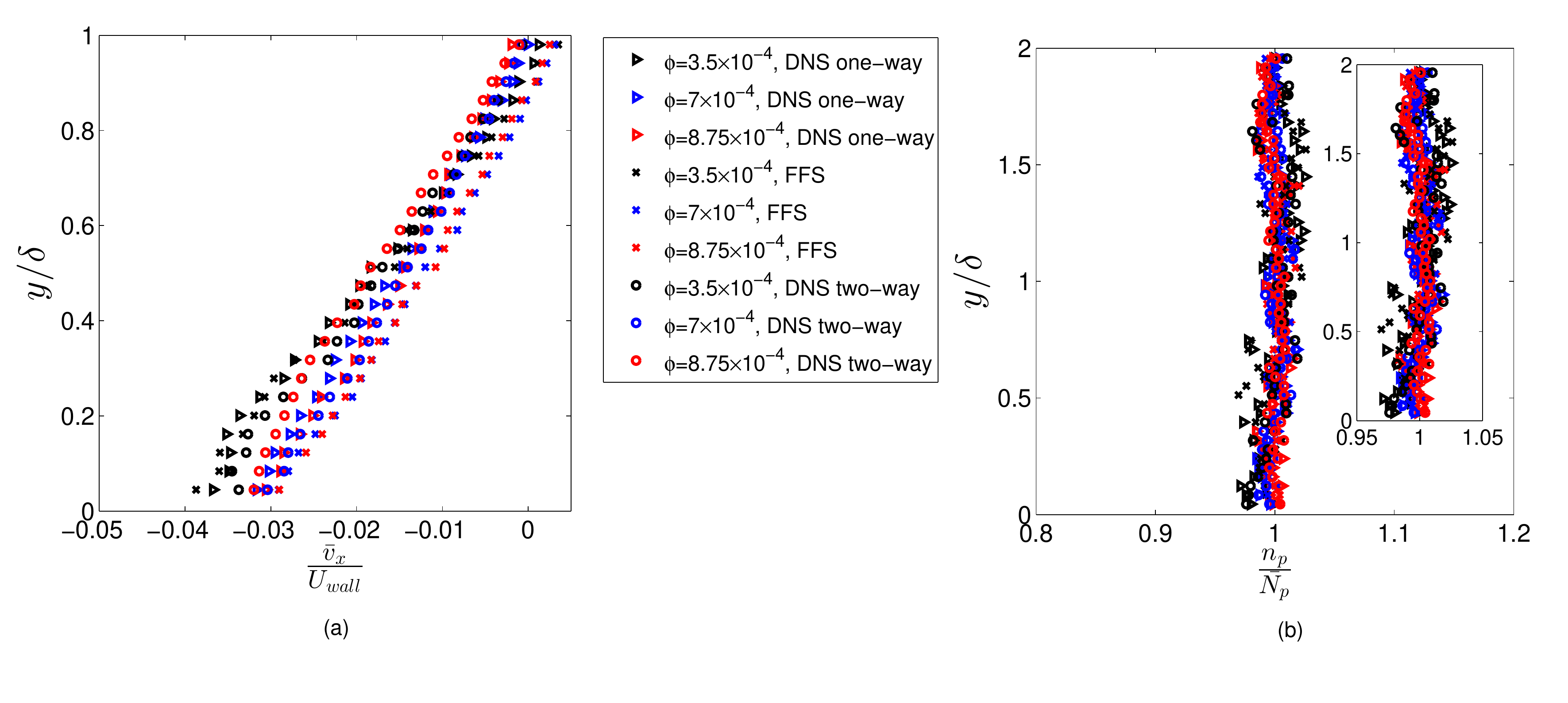}
			\caption{Comparison of (a) mean velocity, and (b) particle number concentration of the particle phase among one-way DNS, two-way DNS and FFS model at different volume fractions} 
			\label{fig:particle_phase_stat1_1way_2way_ffs}
\end{figure*}
\begin{figure*}[!]
			\includegraphics[width=1.0\linewidth]{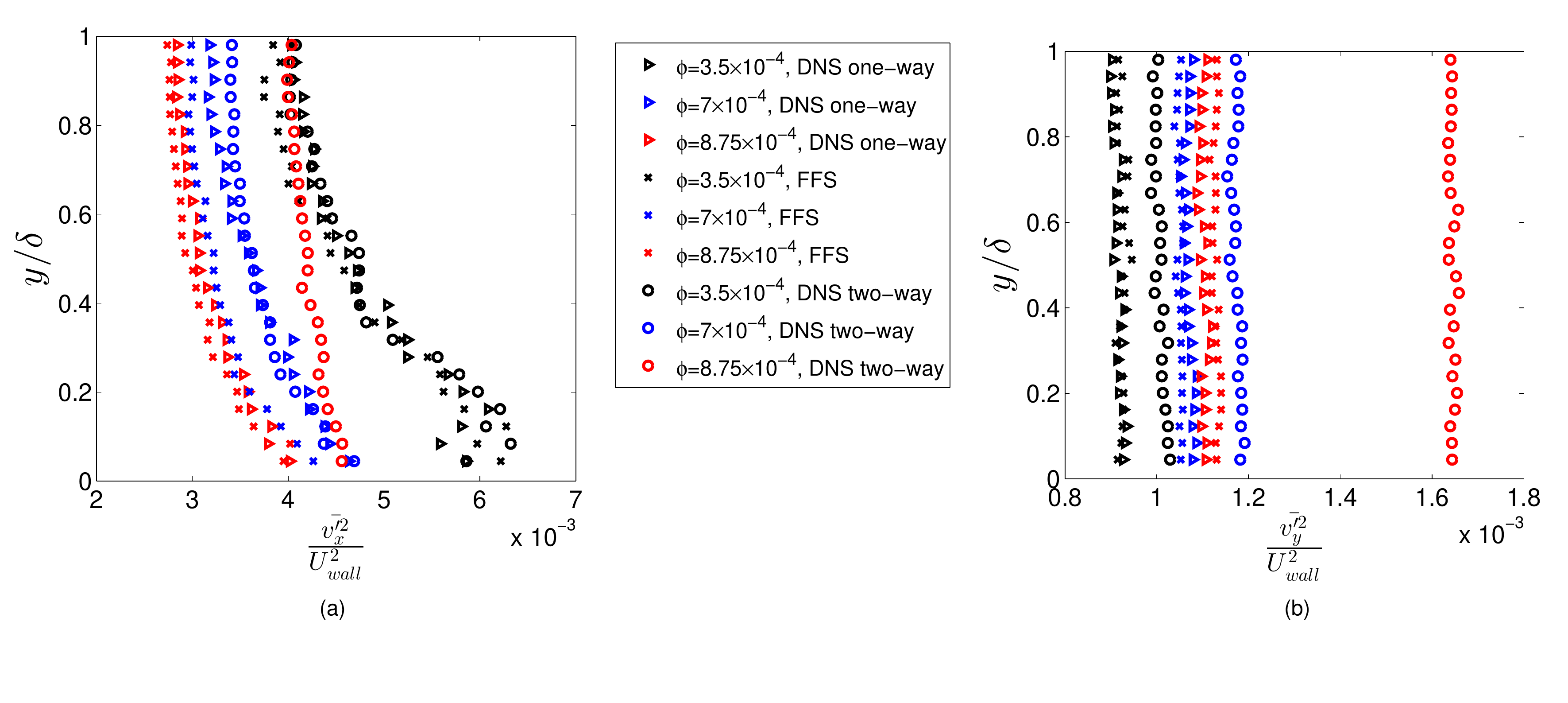}
			\caption{Comparison of (a) streamwise and (b) cross-stream mean square velocity of the particle phase among one-way DNS, two-way DNS and FFS model at different volume fractions}
			\label{fig:particle_phase_stat2_1way_2way_ffs} 
\end{figure*} 
\begin{figure*}[!]
\includegraphics[width=1.0\linewidth]{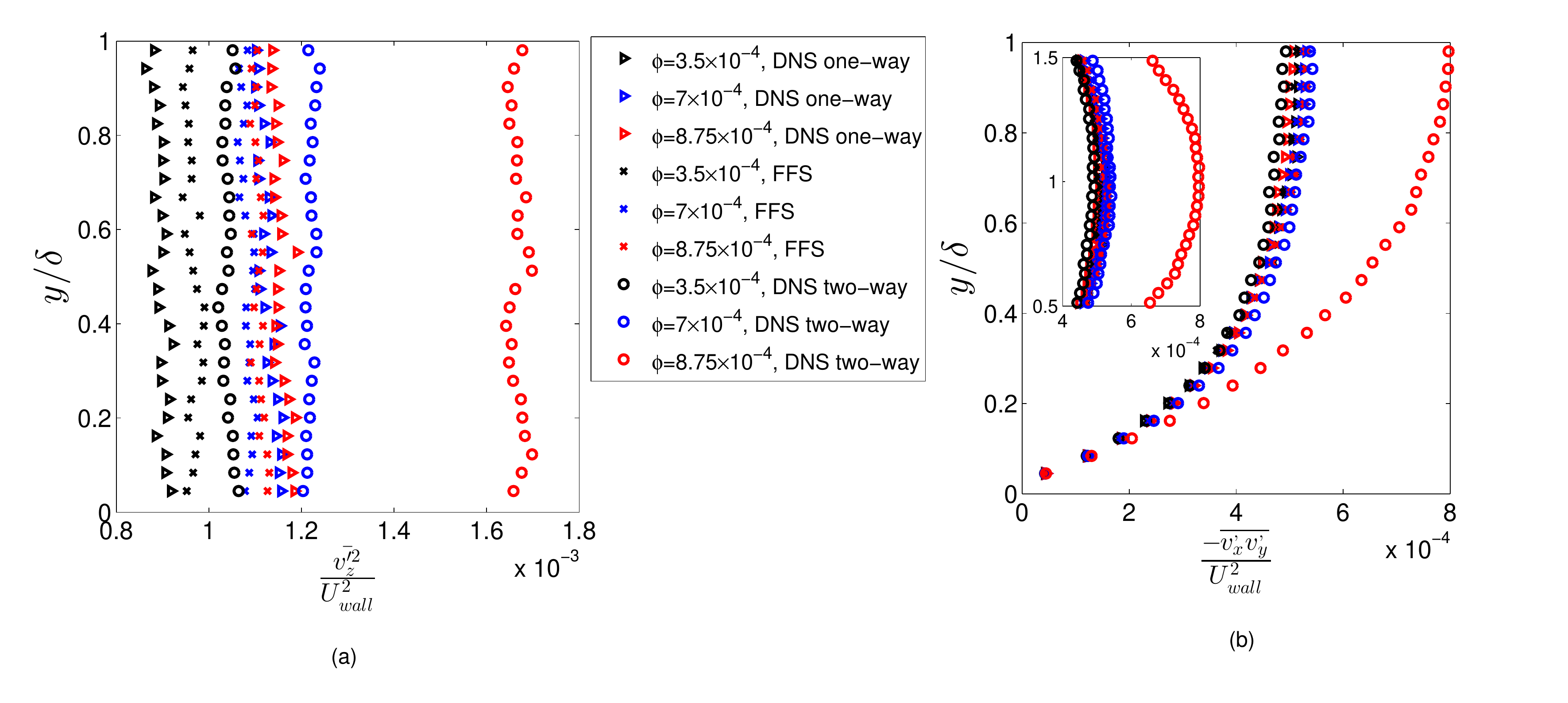}
			\caption{Comparison of (a) spanwise mean square velocity, (b) streamwise and cross-stream velocity cross-correlation of the particle phase among one-way DNS, two-way DNS and FFS model at different volume fractions} 
			\label{fig:particle_phase_stat3_1way_2way_ffs}
\end{figure*}
\begin{figure*}[!]
			\includegraphics[width=1.0\linewidth]{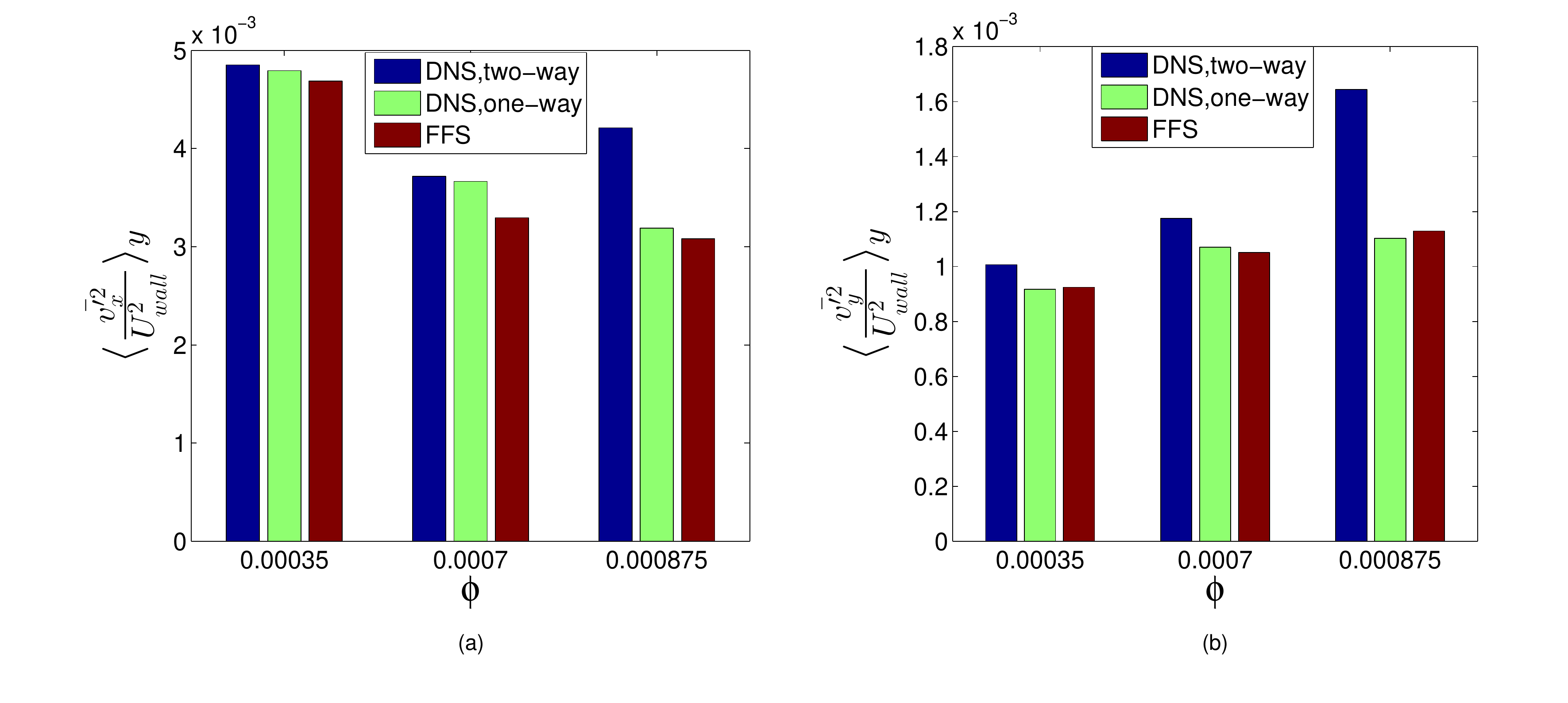}
			\caption{Comparison of spatial average of ($\langle\cdot\rangle_y$) (a) streamwise and (b) cross-stream mean square velocity of the particle phase among one-way DNS, two-way DNS and FFS model at different volume fractions}
			\label{fig:spat_av_particle_phase_stat2_1way_2way_ffs} 
\end{figure*} 
\begin{figure*}[!]
\includegraphics[width=1.0\linewidth]{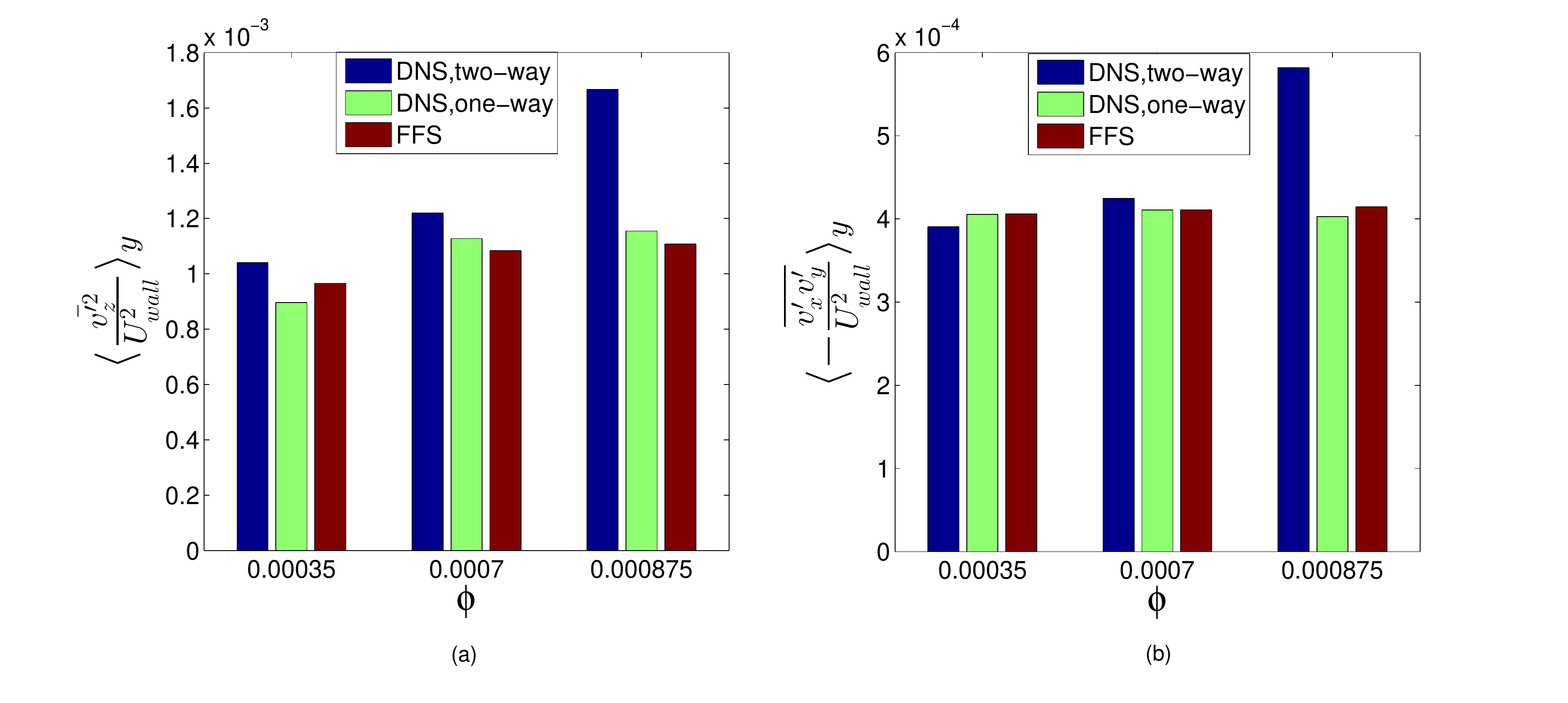}
			\caption{Comparison of spatial average of ($\langle\cdot\rangle_y$) (a) spanwise mean square velocity, (b) streamwise and cross-stream velocity cross-correlation of the particle phase among one-way DNS, two-way DNS and FFS model at different volume fractions} 
			\label{fig:spat_av_particle_phase_stat3_1way_2way_ffs}
\end{figure*}

\textcolor{black}{Figures \ref{fig:particle_phase_stat1_1way_2way_ffs} (a) and (b) shows the mean velocity profile and the concentration profiles for the different volume fraction. It is observed that one-way DNS and FFS results show little deviation from the two-way coupled DNS results for all volume fractions. The second moments of the velocity fluctuations are shown in Figs. \ref{fig:particle_phase_stat2_1way_2way_ffs} and \ref{fig:particle_phase_stat3_1way_2way_ffs}. The figures depict that FFS results quantitatively matches with one-way DNS prediction and show a marginal deviation from the two-way prediction when $\phi<\phi_{cr}$. However, FFS predictions differ significantly from two-way coupled DNS prediction for $\phi>\phi_{cr}$. In this regime FFS overpredicts the second moment of streamwise fluctuation but underpredicts the other components. The average values of the second moments (averaged over half-width) are shown Figs \ref{fig:spat_av_particle_phase_stat2_1way_2way_ffs} and \ref{fig:spat_av_particle_phase_stat3_1way_2way_ffs}. All the components of second moments show a significant error in the FFS predictions of particle phase in the regime where turbulence collapses. The above analysis suggests that the significantly erroneous prediction of FFS beyond the collapse of turbulence may originate either because of the unmodified mean gas velocity or consideration of turbulence diffusivity even after the collapse of turbulence. It is worth noting that FFS does not consider the modification of fluid phase mean velocity and turbulence; rather it uses those properties from the unladen flows. A detailed analysis on the source of the error is discussed in the following sections. First, we perform FFS without turbulent diffusivity beyond the critical volume loading. Next, we explore the effect of modification of mean fluid profile due to higher particle loading. }
\subsection{FFS predictions above critical volume loading in absence of velocity-space diffusivity}
\label{sec:zero_diff_FFS}
In \citet{ghosh2022part1} it is discussed, in details, that the drastic attenuation of turbulence is reflected in the drastic decrease \textcolor{black}{in second moments of fluid velocity fluctuations} along with a modification in the fluid mean velocity profile. Hence for suspensions, with $\phi>\phi_{cr}$, FFS runs with zero turbulent diffusivity  (referred to as zero-diffusivity FFS henceforth) can mimic the actual system approximately. The results discussed in this section is for the volume fraction $\phi=8.75\times10^{-4}$.    
\begin{figure*}[!]
\includegraphics[width=1.0\textwidth]{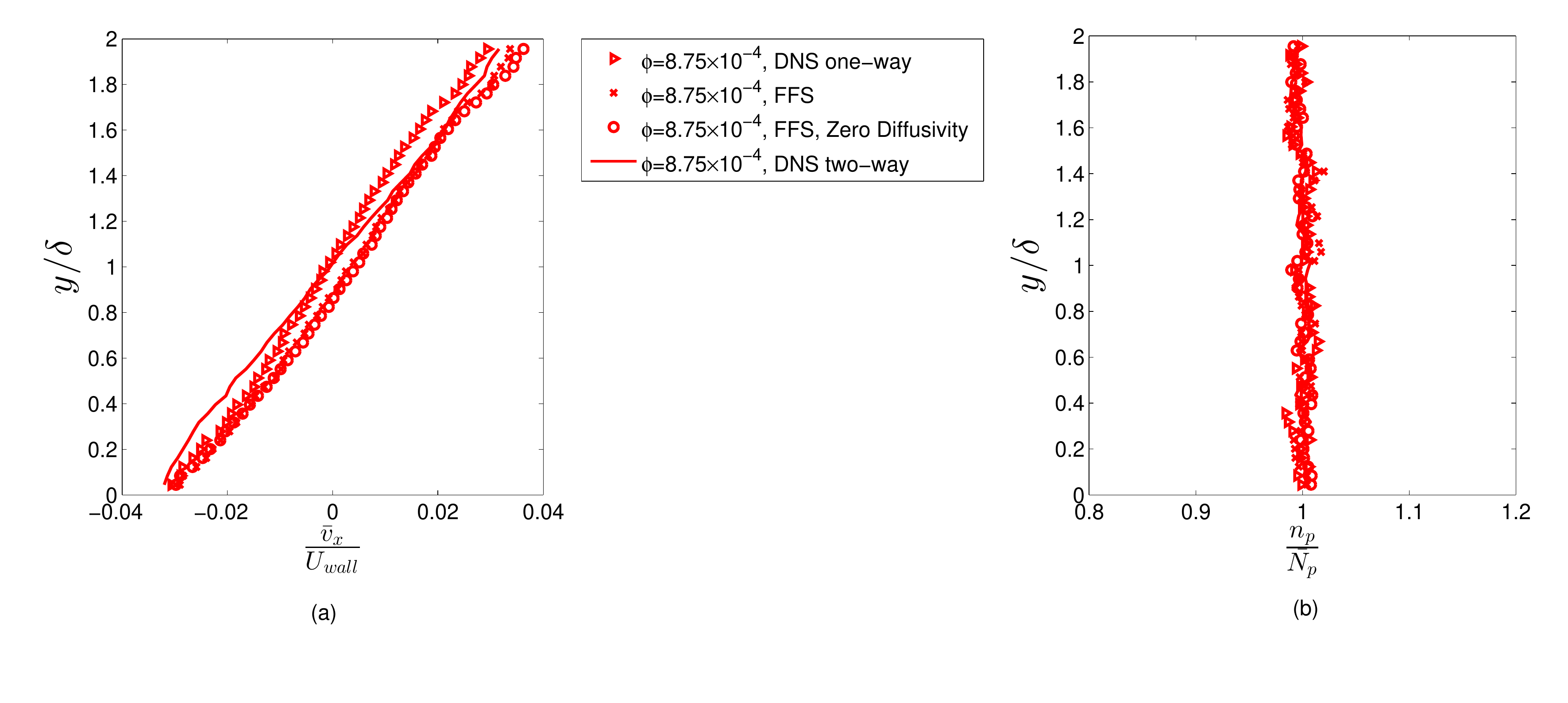}
			\caption{Comparison of (a) mean velocity, (b)  number concentration of particle phase obtained from one-way DNS, two-way DNS and FFS model in absence of velocity-space diffusivity for $\phi=8.75\times10^{-4}>\phi_{cr}$} 
			\label{fig:particle_phase_zero_diff_FFS_1}
\end{figure*}
\begin{figure*}[!]
			\includegraphics[width=1.0\textwidth]{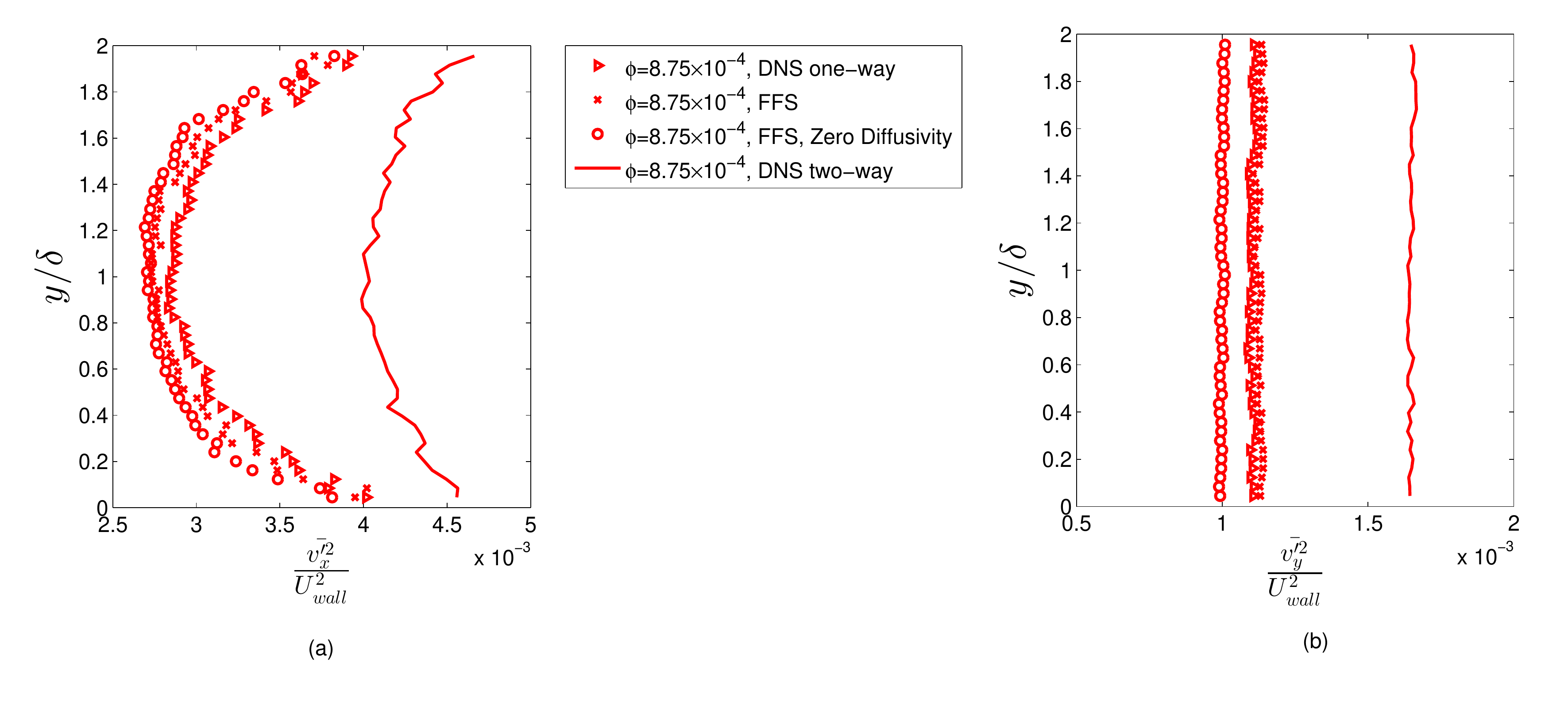}
			\caption{Comparison of (a) streamwise mean square velocity, (b)  cross-stream mean square velocity of particle phase obtained from one-way DNS, two-way DNS, FFS and FFS model in absence of velocity-space diffusivity for $\phi=8.75\times10^{-4}>\phi_{cr}$}
			\label{fig:particle_phase_zero_diff_FFS_2}
\end{figure*} 
\begin{figure*}[!]
\includegraphics[width=1.0\textwidth]{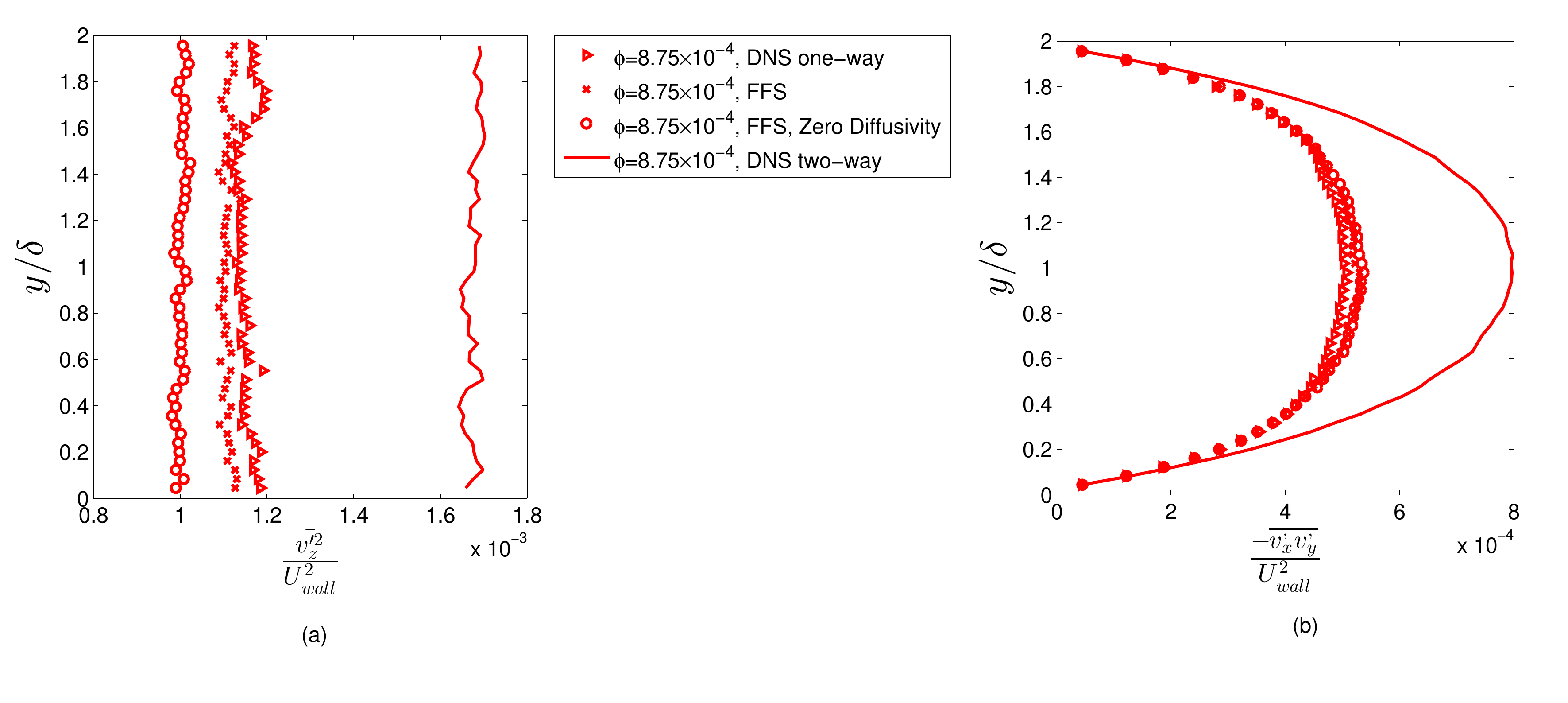}
			\caption{Comparison of (a) spanwise mean square velocity, (b) streamwise and cross-stream velocity cross-correlation of the particle phase obtained from one-way DNS, two-way DNS, FFS and FFS model in absence of velocity-space diffusivity for $\phi=8.75\times10^{-4}>\phi_{cr}$} 
			\label{fig:particle_phase_zero_diff_FFS_3}
\end{figure*}
Figures \ref{fig:particle_phase_zero_diff_FFS_1}-\ref{fig:particle_phase_zero_diff_FFS_3} show a comparative study of the mean particle velocity, number concentration and velocity second moments obtained from zero-diffusivity FFS, FFS, one-way coupled DNS and two-way coupled DNS. Mean velocity predictions by zero-diffusivity FFS remains similar to the other runs as shown in Fig. \ref{fig:particle_phase_zero_diff_FFS_1}. However, Fig. \ref{fig:particle_phase_zero_diff_FFS_2} and \ref{fig:particle_phase_zero_diff_FFS_3} show that the second moment of particle-phase velocity predicted by zero-diffusivity are similar to FFS and one-way coupled DNS and deviates from actual two-way coupled DNS predictions. Hence the absence of fluid phase velocity fluctuations are not solely responsible for the limitation of FFS model in predicting the second moments of particle velocity.
\textcolor{black}{Such an observation motivates to look into the modification of mean fluid velocity profile once the fluctuation collapses.}
\subsection{Modification of FFS (M-FFS)}
The zero-diffusivity FFS runs in section \ref{sec:zero_diff_FFS} shows that the limitation of FFS in predicting the the second moments of particle velocity above $\phi_{cr}$ is not solely due to the drastic decrease of fluid velocity fluctuations. Hence we shift our focus to fluid mean velocity statistics and investigate the change in fluid mean velocity due to turbulence attenuation.  
\begin{figure}[!]
 \includegraphics[width=0.6\linewidth]{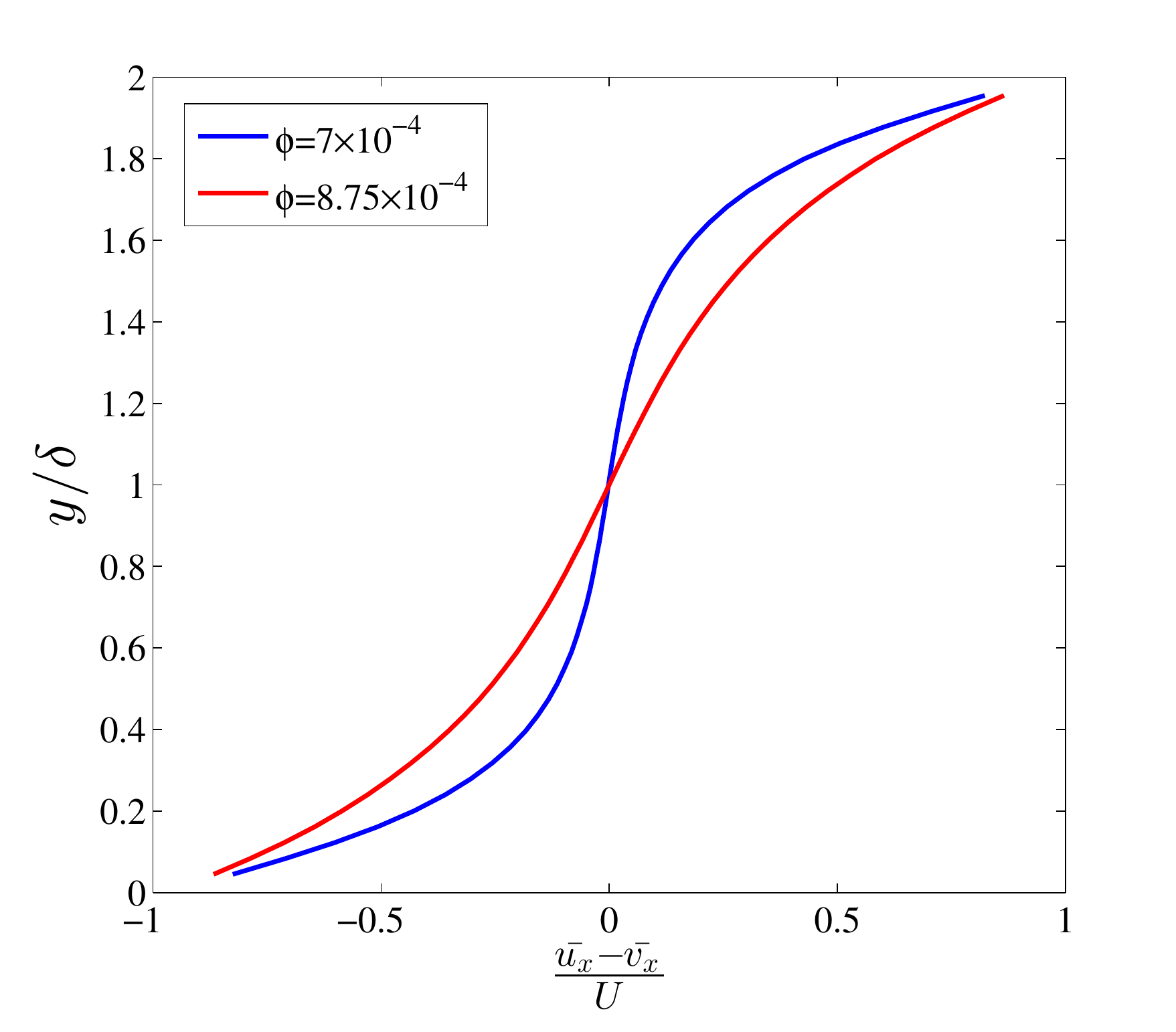}
 	\caption*{(a)}    
\includegraphics[width=0.6\linewidth]{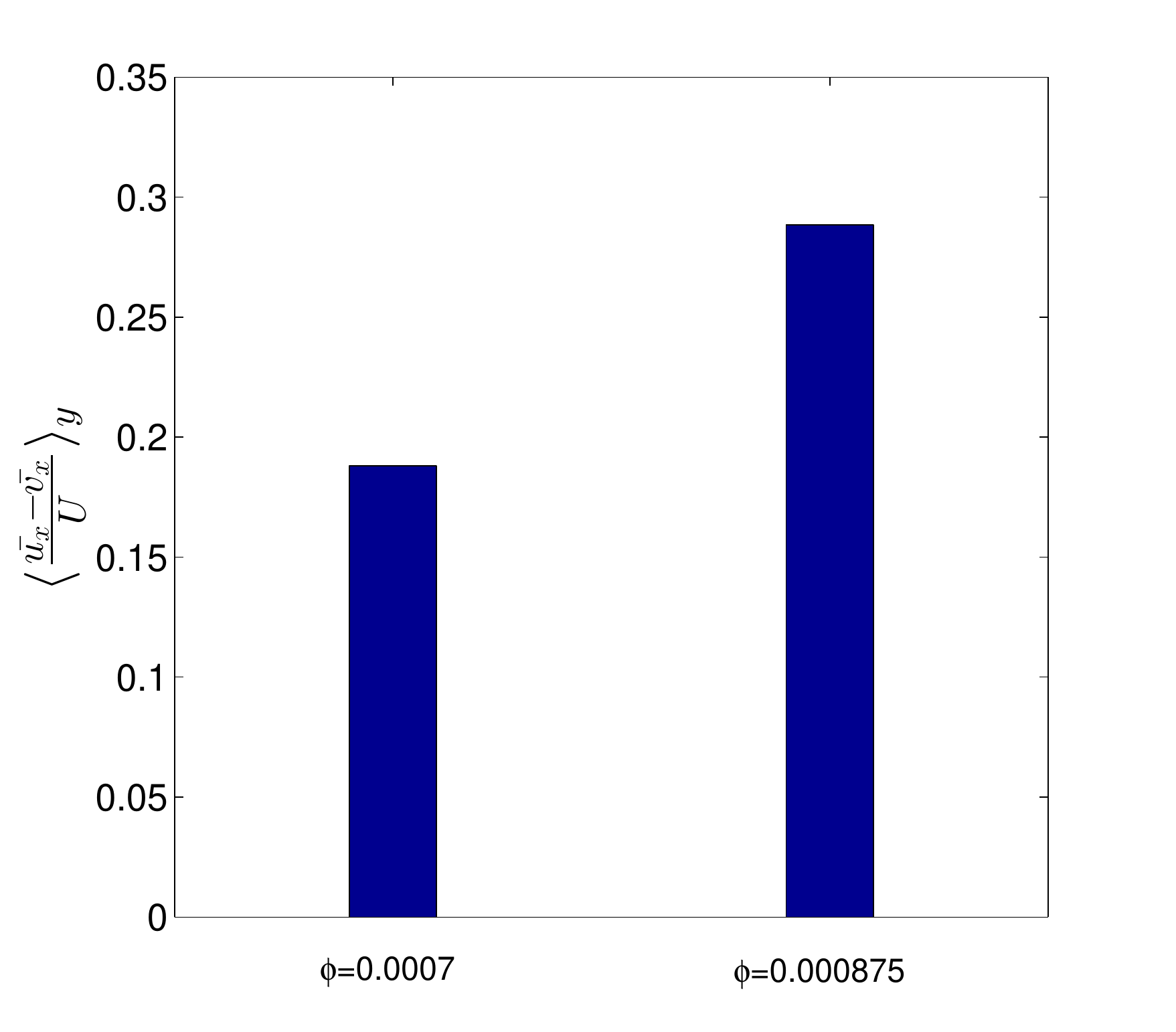}
 	\caption*{(b)}     
\caption{The change in mean velocity difference between fluid-phase and particle-phase before and after turbulence attenuation; (a) profile of mean velocity difference, (b) mean velocity difference averaged over half-width}
\label{fig:fluid_part_vel_diff}
\end{figure}
The mean velocity difference between the fluid-phase and the particle-phase, obtained from two-way coupled DNS, is represented in fig. \ref{fig:fluid_part_vel_diff} as a function of $y$ in Fig. \ref{fig:fluid_part_vel_diff} (a) and as a spatially averaged (over the half-width) quantity in the bar chart Fig.\ref{fig:fluid_part_vel_diff} (b). The change in the trend is reflected in the spatially-averaged quantity which increases ($\sim50\%$) after the turbulence attenuation. This in turn \textcolor{black}{will increase the reverse drag force on the fluid by the particle and will modify viscous stress in the absence of Reynolds stress.} 
\subsubsection{\textbf{Derivation of Modified Fluid Mean velocity Profile}}
\label{sec:derivation_mean_vel}
The mean momentum budget equation of the fluid phase can be written as:
\begin{equation}
   	-\frac{\partial\bar{p}}{\partial x }+\eta_f\frac{d^2\bar{u_x}}{dy^2}+\frac{d}{dy}(-\rho_f\overline{u'_x u'_y})-\overline{\mu_fn_p(u_x-v_x)}=0
   	\label{eq1}
  \end{equation}
  For turbulent Couette-flow 
  \begin{math}
  -\frac{\partial\bar{p}}{\partial x }=0
  \end{math}
  \\Also, for $\phi=\phi_{cr}$ following figure 6 of \citet{ghosh2022part1} ; the transport of Reynolds' stress term decreases by $O(10^{2})$.   
  \begin{math}
  \frac{d}{dy}(-\rho_f\overline{u'_x u'_y})\approx0
  \end{math}
  Equation \ref{eq1} takes the form of
  \begin{equation}
   	\eta_f\frac{d^2\bar{u_x}}{dy^2}-\overline{\mu_fn_p(u_x-v_x)}=0
   	\label{eq2}
  \end{equation}
  Figure \ref{fig:particle_phase_stat_1}(b) shows that the particle number density is constant across $y/\delta$. Therefore, 
  \begin{math}
  n_p\neq f(y)
  \end{math}
   \\ Also following \ref{fig:particle_phase_stat_1}(a), the mean particle velocity varies linearly with $y$ we can write
  \begin{math}
  \bar{v_x}=py+q
  \end{math}
  where $p\approx0.033$ and $q\approx-0.033$.
  Hence from Eq. \ref{eq2}
  \begin{equation} \nonumber
   	\frac{d^2\bar{u_x}}{dy^2}-k(\bar{u_x}-\bar{v_x})=0
  \end{equation}, 
  Where, $k=\frac{\mu_fn_p}{\eta_f}$ and
 \\ \begin{equation}
   	\frac{d^2\bar{u_x}}{dy^2}-k\bar{u_x}=-k\bar{v_x}=-kpy-kq
   	\label{eq3}
    \end{equation}
    The general solution of the above equation is given by:
    \begin{equation}\nonumber
        \bar{u_x}=C_1 exp(y\sqrt{k})+C_2 exp(-y\sqrt{k})
    \end{equation}
    Considering the particular integral as well the solution takes the form :
    \begin{equation}
        \bar{u_x}=C_1 exp(y\sqrt{k})+C_2 exp(-y\sqrt{k})+py+q
        \label{eq4}
    \end{equation}
    k can also be written in terms of volume fraction $\phi$ as: \\
    \begin{math}
    k=18\frac{\phi}{d_p^2}
    \end{math} or $\sqrt{k}=82.7\sqrt{\phi}$, as $d_p=0.0513$.
   \\ The boundary conditions for equation \ref{eq4} are:
    $\bar{u_x}=-1$ at $y=0$ and $\bar{u_x}=0$ at $y=1$.
    By putting the values of $p$ and $q$, the constants $C_1$ and $C_2$ come out as follows:
    \begin{equation}
    C_1=0.967\frac{exp(-165.4\sqrt{\phi})}{1-exp(-165.4\sqrt{\phi})}
    \end{equation}
    \begin{equation}
    C_2=-\frac{0.967}{1-exp(-165.4\sqrt{\phi})}    
    \end{equation}
    Hence the solution of mean fluid velocity for $\phi>\phi_{cr}$ can be predicted using the solution, represented as a function of $\phi$, given below:
    \begin{eqnarray}
     \bar{u_x}&=0.033y +0.967\frac{exp(-165.4\sqrt{\phi})}{1-exp(-165.4\sqrt{\phi})}exp(82.7y\sqrt{\phi})\\&-\frac{0.967}{1-exp(-165.4\sqrt{\phi})}exp(-82.7y\sqrt{\phi})-0.033
     \label{eq:mean_vel}
    \end{eqnarray}
    \begin{figure}
        \centering
        \includegraphics[width=0.5\textwidth]{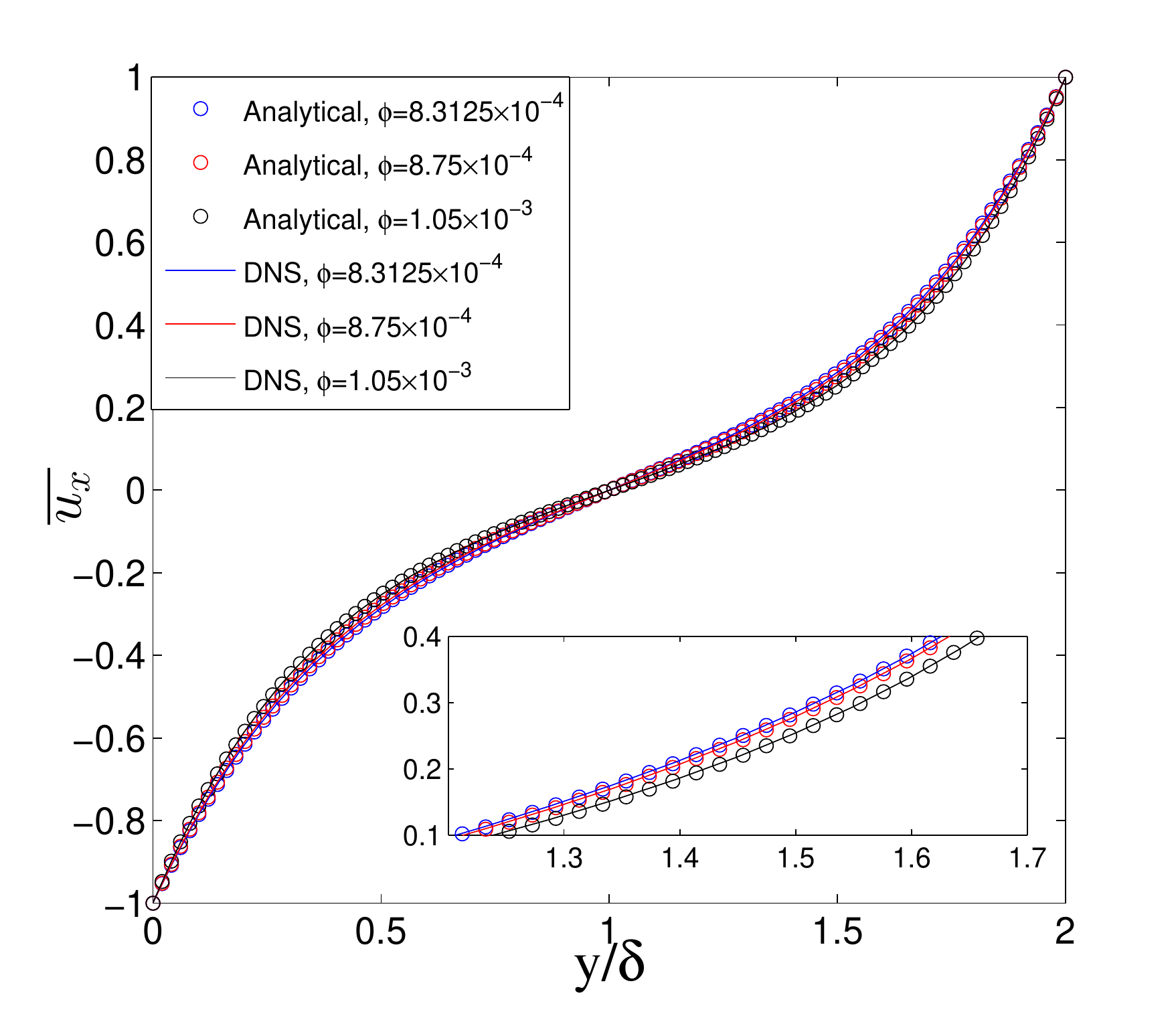}
        \caption{Comparison of mean fluid velocity obtained analytically from eq.\ref{eq:mean_vel} and from two-way coupled DNS data for $\phi>\phi_{cr}$}
        \label{fig:DNS_analytical}
    \end{figure}
    Figure \ref{fig:DNS_analytical} shows the comparison of the mean fluid velocities for $\phi>\phi_{cr}$ obtained from Eq.\ref{eq:mean_vel} and two-way coupled DNS. The near-perfect collapse portrays the accuracy of the mean-velocity estimate from Eq. \ref{eq:mean_vel}. \textcolor{black}{We include the eq. \ref{eq:mean_vel} in FFS simulation and use this modified FFS (M-FFS) technique to predict the particle phase dynamics.}     
\subsubsection{\textbf{Particle Velocity Statistics Using M-FFS}}
The formulation for estimating the modified \textcolor{black}{FFS (M-FFS)} is discussed in the previous subsection \ref{sec:derivation_mean_vel}. \textcolor{black}{Here we compare the M-FFS prediction with two-way coupled DNS results.}
This subsection comprises of comparing the simulation data for $\phi=8.75\times10^{-4}>\phi_{cr}$, obtained from the following three cases:
(a) two-way coupled DNS, (b) FFS,  and (c) M-FFS. 
\begin{figure*}[!]
\includegraphics[width=1.0\textwidth]{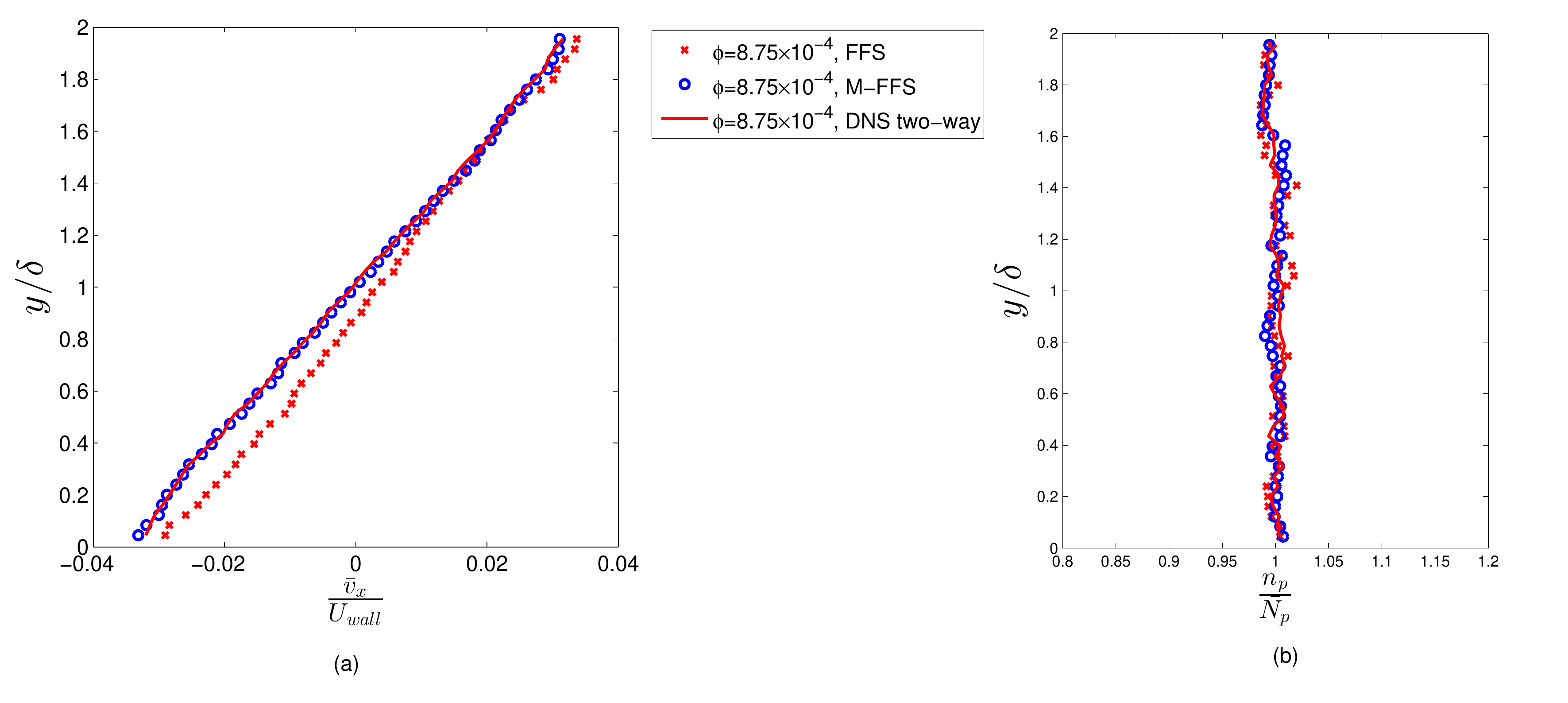}
			\caption{Comparison of (a) mean velocity, (b)  number concentration of particle phase obtained from two-way DNS,  FFS and M-FFS model for $\phi=8.75\times10^{-4}>\phi_{cr}$} 
			\label{fig:mod_particle_phase_zero_diff_FFS_1}
\end{figure*}
\begin{figure*}[!]
			\includegraphics[width=1.0\textwidth]{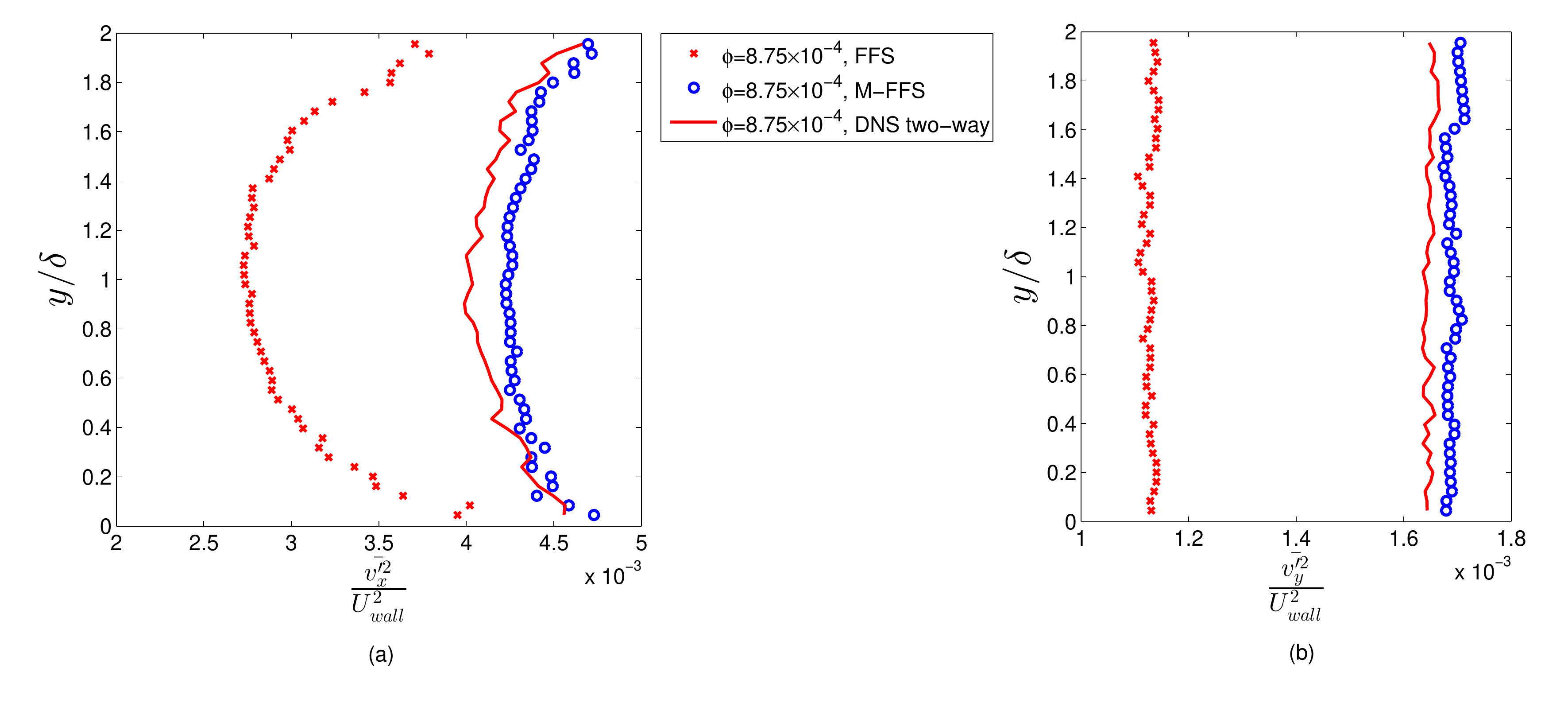}
			\caption{Comparison of (a) streamwise mean square velocity, (b)  cross-stream mean square velocity of particle phase obtained from two-way DNS,  FFS and M-FFS model for $\phi=8.75\times10^{-4}>\phi_{cr}$}
			\label{fig:mod_particle_phase_zero_diff_FFS_2}
\end{figure*} 
\begin{figure*}[!]
\includegraphics[width=1.0\textwidth]{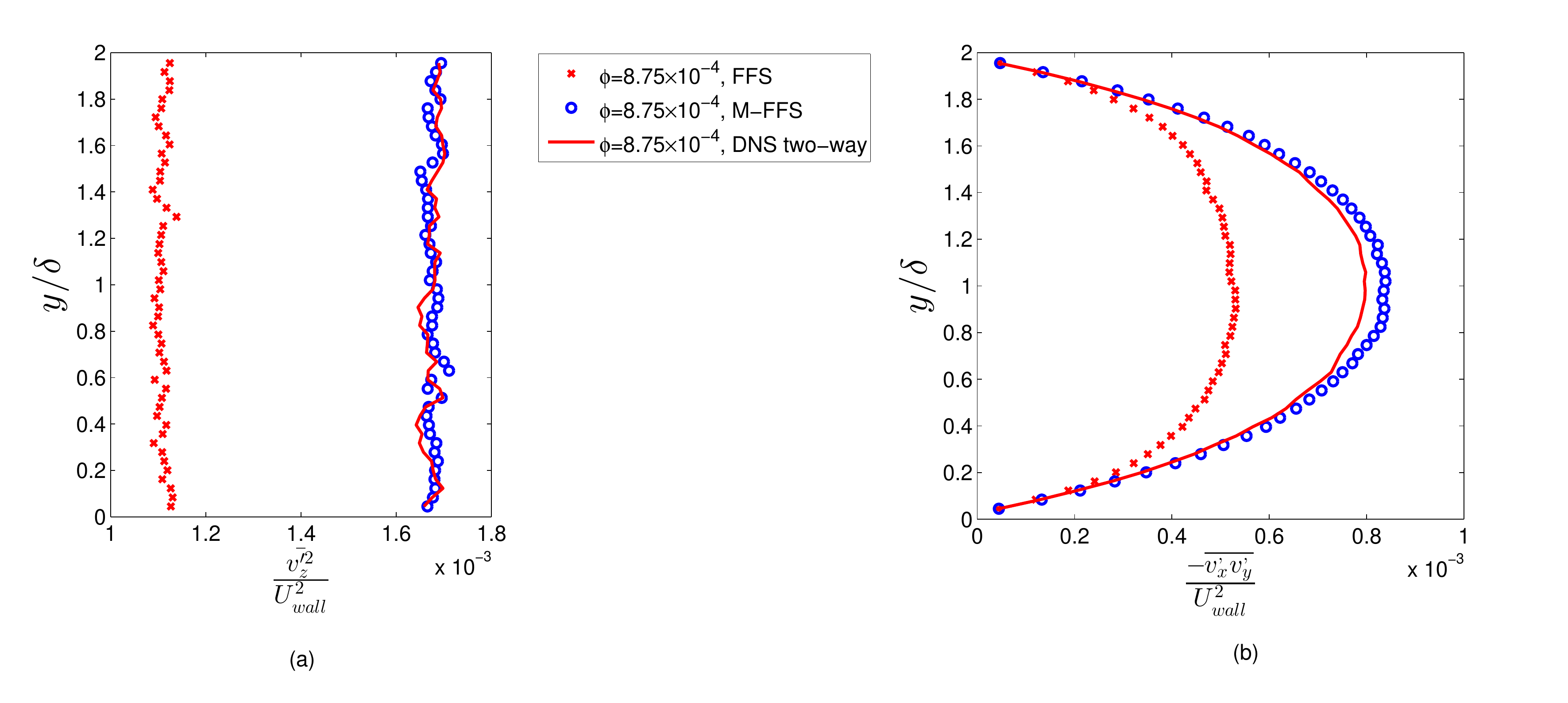}
			\caption{Comparison of (a) spanwise mean square velocity, (b) streamwise and cross-stream velocity cross-correlation of the particle phase obtained from two-way DNS,  FFS and M-FFS model for $\phi=8.75\times10^{-4}>\phi_{cr}$} 
			\label{fig:mod_particle_phase_zero_diff_FFS_3}
\end{figure*}
Figure \ref{fig:mod_particle_phase_zero_diff_FFS_1} depicts the mean particle velocity profile and the number concentration profile. The M-FFS  does not show any deviation from the rest of the cases (a) and (b). However, the second moments of particle velocities, shown in Fig. \ref{fig:mod_particle_phase_zero_diff_FFS_2} and \ref{fig:mod_particle_phase_zero_diff_FFS_3}, show the high accuracy of M-FFS runs (case (c)) with respect to the two-way coupled DNS data. Hence it can be safely substantiated that M-FFS  can predict the particle velocity statistics for suspensions denser than the volume fraction $\phi_{cr}$, where the effect of reverse feed-back force \textcolor{black}{on the fluid phase is balanced with the divergence of viscous stress.}  
\section{Conclusion}
This article focuses on the particle phase statistics of particle-laden sheared turbulent suspension, in presence of ideal, elastic, smooth inter-particle and wall-particle collisions. In the first section \ref{sec:Particle Phase Statistics ideal}, it is observed that due to the high volume fraction of the heavy inertial particles, the particles distribute uniformly across the channel. Mean particle velocity profile becomes marginally flatter with the increase in volume loading. 
\textcolor{black}{With increase in volume fraction streamwise particle velocity fluctuation decreases and profile becomes more flat.} The mean-square velocity profiles in cross-stream and spanwise direction are flat and increase with volume fraction due to the dominance of collision generated stresses. However in the regime, where turbulence attenuation has taken place, the $\overline{v_y'^2}$, $\overline{v_z'^2}$ and $\overline{v_x'v_y'}$ deviates significantly from the regime before attenuation.
The range of volume factions \textcolor{black}{used in the simulations} belong to the regime $\tau_{cp-w} < \tau_{cp-p} < \tau_v $ (except for $\phi=1.75\times10^{-4}$ where $\tau_{cp-w} < \tau_v< \tau_{cp-p} $). Hence collisional dynamics, especially  wall-particle collisions, play a more important role than the viscous relaxation \textcolor{black}{mechanism} in the particle phase dynamics. 
The effect of reverse feed-back force on particle phase velocity fluctuations is marginal for the suspension having volume fractions less than $\phi_{cr}$. However, this effect becomes significant \textcolor{black}{when the volume fraction is} above critical volume fraction.
A comparison of FFS model predictions with one-way and two-coupled DNS results \textcolor{black}{shows that} FFS predictions \textcolor{black}{compare well} with one-way DNS results \textcolor{black}{and marginally deviates from two-way DNS results}. This indeed extends the applicability of the FFS model in the volume fraction parameter-space up to critical volume fraction for high inertial particles. However, FFS results and one-way coupled DNS results show significant deviation for $\phi=(8.75\times10^{-4})>\phi_{cr}$, where the turbulence \textcolor{black}{modulates drastically}. In this regime, FFS with modified mean fluid velocity profile and zero-diffusivity can successfully predict the particle phase statistics. A detailed formulation is shown where the mean fluid velocity can be expressed as a function of volume fraction. Hence it is established here that the particle phase statistics of high inertial particles can also be predicted with less expensive decoupled \textcolor{black}{M-FFS (modified FFS)} model in place of two-way coupled DNS, in denser volume fraction regime, if an a priori knowledge of $\phi_{cr}$ becomes available.

		\begin{acknowledgments}
				We wish to acknowledge the financial support of SERB, DST, Government of INDIA.
			\end{acknowledgments}
		
		\section*{Data Availability Statement}
		The data that support the findings of this study are available
		from the corresponding author upon reasonable request.

\bibliography{bib_joined.bib}	
\end{document}